\documentclass{article}
\PassOptionsToPackage{table}{xcolor}
\usepackage{iclr2027_with_preprint,times}
\usepackage{microtype}\usepackage[normalem]{ulem}
\usepackage[T1]{fontenc}
\usepackage[table]{xcolor}
\definecolor{CornflowerBlue}{HTML}{6495ED}
\usepackage{float}
\usepackage[colorlinks=true,
            linkcolor=CornflowerBlue,
            citecolor=CornflowerBlue,
            urlcolor=CornflowerBlue]
            {hyperref}
\usepackage{amsmath}
\usepackage{graphicx}
\usepackage{cleveref}
\usepackage{booktabs}
\usepackage{array}
\usepackage{tabularx}
\usepackage{listings}
\usepackage{placeins}
\usepackage{microtype}\usepackage[normalem]{ulem}
\newcolumntype{L}[1]{>{\raggedright\arraybackslash}p{#1}}
\newcolumntype{Y}{>{\raggedright\arraybackslash}X}
\definecolor{codekeyword}{RGB}{126,46,141}
\definecolor{codestring}{RGB}{3,106,7}
\definecolor{codespecial}{RGB}{0,92,168}
\lstdefinestyle{quirkpython}{
  language=Python,
  basicstyle=\ttfamily\fontsize{8}{10}\selectfont,
  keywordstyle=\color{codekeyword}\bfseries,
  stringstyle=\color{codestring},
  emph={w_value,__import__,final_answer,__call__,step_14,SimpleNamespace},
  emphstyle=\color{codespecial},
  showstringspaces=false,
  columns=fullflexible,
  keepspaces=true
}
\usepackage{float}
\graphicspath{{figures/}}
\hypersetup{
  colorlinks=true,
  linkcolor=CornflowerBlue,
  citecolor=CornflowerBlue,
  urlcolor=CornflowerBlue,
  pdftitle={Surprise Masking: Private Data Filtering with Model Likelihoods},
  pdfauthor={Riya Tyagi, Sahra Ghalebikesabi, Vinnie Monaco, Avi Schwarzschild, Om Thakkar},
  pdfdisplaydoctitle=true
}

\title{Surprise Masking: Private Data Filtering with Model Likelihoods}

\author{Riya Tyagi\thanks{Work done while the author was an intern at OpenAI.}, Sahra Ghalebikesabi, Vinnie Monaco, Avi Schwarzschild, Om Thakkar \\
OpenAI \\
\texttt{\{rtyagi,sahra,vinnie,avis,om\}@openai.com}
}

\iclrpreprint

\begin{document}
\maketitle

\begin{abstract}
Data useful for language model (LM) training can contain private or unwanted information that is at risk of being learned or memorized.
We propose Surprise Masking (SM), a simple form of perplexity filtering that uses a publicly available reference model to identify unwanted spans and suppress learning from them while preserving the rest of each example.
On synthetic code benchmarks, we show SM improves the privacy-utility tradeoff over standard masking methods and PII filtering.
In fine-tuning experiments, SM reduces approximate extraction of private values from revealed prefixes from 66.3\% to 2.3\% and the reproduction of stylistic coding quirks from 23.6\% to 1.3\%, both while preserving performance on the evaluated coding tasks.
We also evaluate SM for sanitizing datasets by directly removing unwanted content.
Here, SM outperforms DePA, a perplexity-based defense for detecting poisoned dead code, while specialized detectors perform better on broader credential and PII benchmarks.
Across eleven settings, we find that SM's downstream success is predictable from the separation of unwanted and useful content in the distribution, and it is particularly useful with code data.
Ultimately, SM is an inexpensive, adaptable method to preserve valuable examples while reducing the learning of unwanted information.
\end{abstract}

\section{Introduction}
\label{sec:intro}

\begin{figure}[b]
    \centering
    \includegraphics[width=\textwidth]{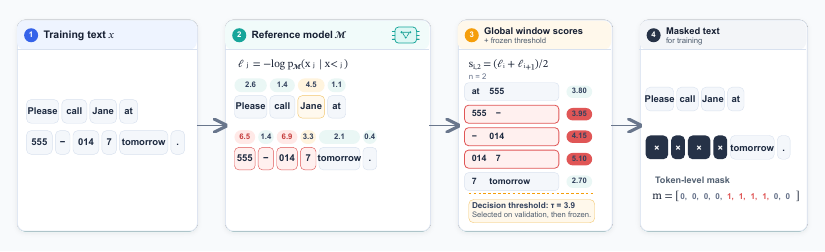}
    \caption{Overview of Surprise Masking, illustrated with Global SM. A reference model scores token spans, and sufficiently surprising spans are masked from the training loss.}
    \label{fig:example_image}
\end{figure}

Training samples that are valuable for learning can also contain undesirable content, including private information.
This leaves practitioners with limited options. They can discard these samples, reducing leakage risk but sacrificing useful information, or train on them with safeguards that have their own costs and limitations.
Alternatively, training-time interventions offer a way to alter how models learn from risky data~\citep{hans2024likegoldfishdontmemorize, wichers2025inoculate,tan2025inoculate}.
These methods tend not to offer explicit or theoretical guarantees, so their privacy protections need empirical testing.
Another approach is post-training unlearning methods, which typically localize information within a model before removing it \citep{merullo2025memorizationreasoningspectrumloss,maini2023localized, stoehr2024paragraph}.
These tools can all be costly and have unpredictable effects on model behavior. 
We aim to improve safeguards that mask unwanted portions of data samples before training on them. 

In this work, we revisit \textit{perplexity filtering}, a method that uses a reference language model's perplexity scores to filter data (see \Cref{fig:example_image} for an overview).
Prior work uses perplexity to select useful training data \citep{ankner2024perplexed,marion2023pruning,thrush2025pretrain} and \citet{wu2025stm} use it to mask high-perplexity tokens from the fine-tuning loss to mitigate forgetting. 
We investigate this same principle for suppressing the learning of private and other unwanted content.
Our approach, \textit{Surprise Masking (SM)}, masks token spans that are unlikely under a publicly available reference model.

The intuition is that private content may be unfamiliar to the reference model and therefore surprising, so masking these spans can reduce how much a model trained on the masked data learns about private information. 
Note the difference between \textit{private content} and \textit{sensitive content}: the reference model might know sensitive content, like email addresses appearing in public datasets; however, the presence of such information makes it not private.
SM aims to reduce the leakage of private content, especially high-entropy secrets such as API keys, passwords or account numbers.

We first evaluate SM as a training-time intervention. 
We mask selected tokens from the training loss while retaining the rest of each example. 
In this setting, we evaluate SM on private content by generating synthetic training datasets with sensitive information that is unavailable to the reference model.
We evaluate utility using downstream model performance.
The following two existing observations suggest SM may be particularly effective here.
First, perplexity can sometimes predict memorization when other properties of the data are held constant \citep{meeus2024copyrighttrapslargelanguage}.
Second, false positives--surprising spans that are erroneously masked--might contain poor quality content, such as gibberish, and thus have little effect on downstream model performance.

We additionally evaluate SM as a data sanitization method. 
These benchmark tasks test whether reference model surprise can localize sensitive content while preserving the surrounding data. 
Sanitization might be useful when sensitive information must be removed from data, for instance before releasing a dataset.
Here, we measure privacy by how much sensitive content is removed and utility by how much other content is preserved.

Our results show that during pretraining and fine-tuning, SM extends the privacy--utility frontier beyond existing baselines (see~\Cref{sec:sm-on-code}).
In the sanitization setting, SM remains useful on some narrow distributions. Ultimately, it performs well on mitigating dead code poisoning, but specialized detectors are generally stronger on broader credential and PII benchmarks (see~\Cref{sec:sanitization}). 
In~\Cref{sec:distributional-account}, we characterize when SM succeeds and fails, and show how practitioners can cheaply assess its suitability for a new setting.
Ultimately, we see SM as an inexpensive, adaptable complement to scrubbers, rewriters, and other tools in a robust privacy suite.

\section{Method}
\label{sec:method}

Our method can be described as four steps, each with some runtime choices or hyperparameters.
At the most general level, SM accepts four inputs: a data example $T$, a reference model $M$, a window function $W$, and a threshold function $\tau$.
The window function maps the example to potentially overlapping windows, while the threshold function determines how surprising each window must be to be masked.
SM returns a mask at the window level, while a post-processing step produces a token-level mask.

Formally, let $\mathcal{V}$ denote a finite vocabulary of tokens, and let $T=(t_1,\ldots,t_N)\in\mathcal{V}^N$ be a sequence of tokens.
Let $M$ be an autoregressive language model that serves as our \emph{reference model}.
For privacy applications, we recommend choosing a publicly available $M$ that was not trained on the private content we want to exclude.
We call the content that should be masked \emph{target content}, and refer to the unintended masking of non-target content as \emph{collateral masking}.
The four steps are as follows.
\begin{enumerate}
    \item \textit{Construct windows.} The window function $W$ maps $T$ to a collection of $K$ possibly overlapping \space windows:
    \begin{equation*}
\label{eq:window}
        W(T)=\{w_i\}_{i=1}^{K},
        \qquad
        w_i\subseteq\{1,\ldots,N\}.
    \end{equation*}
    For fixed-width windows of length $n$, $w_i=\{i,\ldots,i+n-1\}$.
    Alternative window functions can return complete code lines, document fields, or chat responses.
    Complete coverage is not required by this definition, leaving the possibility for positions to be outside all windows and receive no window-based mask.

    \item \textit{Compute surprise.} For each window, compute its mean negative log likelihood under the reference model:
    \begin{equation}
    \label{eq:window-surprise}
        L_i
        =
        -\frac{1}{|w_i|}
        \sum_{j\in w_i}
        \log p_M\!\left(t_j\mid t_{<j}\right).
    \end{equation}
    Higher values indicate that the window is more surprising to $M$.

    \item \textit{Compute thresholds.} Assign each window a threshold $\tau_i=\tau(w_i)$.

    \item \textit{Select windows.} Mark windows whose surprise is at or above their threshold, setting $b_i=\mathbf{1}\!\left\{L_i\geq\tau_i\right\}$.
\end{enumerate}

The output of these four steps of SM is a collection of windows and their masking decisions which we can express as $\operatorname{SM}(T;M,W,\tau)
    =
    \{(w_i,b_i)\}_{i=1}^{K}$.
Different choices of $M$, $W$, and $\tau$ can adapt SM to different settings.
\textit{Global SM} uses the same threshold for every window, $\tau_i = \tau_0$, while \textit{Local SM} selects windows by their surprise rank among nearby windows.
\textit{Prompted SM} conditions the reference model on a description of non-target content, which can make target content more surprising.
We use different configurations throughout~\Cref{sec:sm-on-code,sec:sanitization} and provide variant details in~\Cref{sec:appendix-ablations}.

Once SM selects surprising windows, the simplest map to a token-level mask is to mask every token that belongs to at least one selected window.
Let $m_j$ denote the masking of token $j$ where $m_j = 1$ means we mask it from the loss, i.e. we do not backpropagate on its prediction.
\begin{equation}
\label{eq:global-decision}
    m_j
    =
    \mathbf{1}\!\left\{
        \exists i:\,
        j\in w_i
        \ \text{and}\
        b_i=1
    \right\}.
\end{equation}
Additional post-processing options can yield different token masks.
For example,  \textit{Consensus masking} requires a token to appear in at least $k$ selected windows before it is masked, while \textit{duplicate propagation} masks repeated occurrences of selected spans.
Both can be applied individually or together to any SM output.

\subsection{Use cases}
\label{sec:eval-procedures}

We evaluate SM both as a \space training intervention and as a data sanitization approach.
When SM is used as a training-time intervention, masked tokens are excluded from the training loss but remain in the context window, so they still provide context for predicting later tokens.
In this setting, the notion of privacy measures whether training causes the model to learn an unwanted attribute, while utility captures model performance on the intended task.
In our experiments, the unwanted attributes are private values and user-specific stylistic quirks embedded in code, while the intended task is to produce functionally correct code.
Training can be forgiving because collateral masking may not reduce utility when the masked tokens are unimportant for learning the intended task.
We present training-time intervention results in~\Cref{sec:sm-on-code}.

Sanitization, however, is less forgiving.
Here, masked tokens are redacted, so every non-target token removed counts as an error regardless of its downstream training value.
We evaluate SM in sanitization settings covering credentials, personal information, and poisoned code in~\Cref{sec:sanitization}.
We measure privacy by recall (TPR) on target content and utility by preservation of non-target content, which is $1-$FPR.

SM works best as a training intervention and on narrowly defined sanitization tasks, such as detecting poisoned code.
On broader sanitization tasks, it can achieve high recall, but frequent false positives ultimately limit its usefulness.
\Cref{sec:distributional-account} examines when SM works well by comparing the surprise score distributions of target and non-target content.

\section{Surprise Masking suppresses unwanted code attributes while preserving utility}
\label{sec:sm-on-code}

We first evaluate SM as a training-time intervention, asking whether it suppresses private values and stylistic quirks in code while preserving useful training signal.
For fine-tuning, we use GPT-OSS-20B \citep{openai2025gptoss}.
Because it already performs well on common coding benchmarks and our fine-tuning updates are relatively small, utility differences between models trained with and without SM are difficult to measure.
To simplify utility measurement, we build two datasets based on the ExStruct coding library~\citep{harumiweb2026exstruct}, released after GPT-OSS-20B's knowledge cutoff~\citep{openai2025gptoss}; before tuning, the model solves none of our ExStruct tasks, leaving clear room to measure learning.

Both datasets are designed around a realistic model development pipeline in which developers train a model on user code to improve its performance on ExStruct.
In \textit{Code PII}, the user code contains private values; in \textit{Code Quirks}, it exhibits user-specific stylistic habits.
First, we prompt GPT-5.6 to produce realistic ExStruct tasks and matching code solutions. We then modify the resulting code in controlled ways to introduce synthetic private values or stylistic quirks.
These controlled modifications provide ground-truth labels for the target tokens that SM should mask.
For fine-tuning, we turn each task-solution pair into a chat: the task becomes the user request, and the modified code solution becomes the assistant response.

Across both fine-tuning and pretraining, we sweep several SM thresholds $\tau$ to trace SM's privacy--utility curve.
The specific $\tau$ values differ per dataset such that each sweep covers a useful privacy--utility range.
We compare SM with \textit{unmasked training}, an \textit{oracle} that masks exactly the labeled target spans, \textit{random masking} at the same token budget, Goldfish loss ~\citep{hans2024likegoldfishdontmemorize}, which selects tokens using a deterministic hash of their preceding context, and the OpenAI Privacy Filter (OPF), a trained PII scrubber~\citep{openai2026privacyfilter}.
Additional dataset construction and training details appear in \Cref{sec:appendix-datasets,sec:appendix-implementation}.

\begin{figure}[H]
    \centering
    \includegraphics[width=\linewidth]{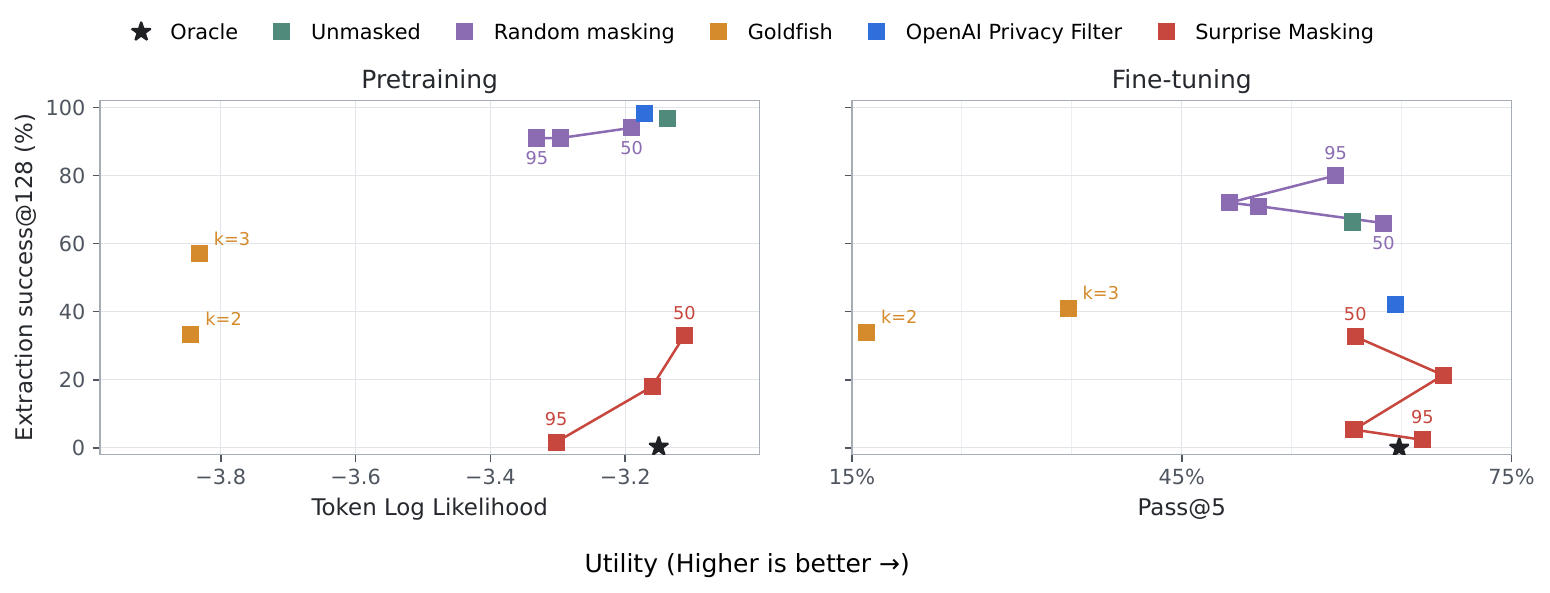}
    \caption{
    Privacy--utility tradeoffs on Code PII.
The vertical axis shows the fraction of private values approximately extracted at least once in 128 samples from a revealed prefix.
The left panel shows ExStruct token log likelihood after pretraining, and the right panel shows ExStruct pass@5 after fine-tuning.
Lower extraction success and higher utility are better.
Global SM uses windows of four tokens, consensus across four windows, and duplicate propagation.
Endpoint labels give SM's intended validation target recall in percent, and the same labels identify random masking at the corresponding token budgets.
Goldfish masks approximately $1/k$ of eligible tokens.
Points show means across training seeds.
}
    \label{fig:pii-main}
\end{figure}

\paragraph{Code PII.}
Our Code PII fine-tuning dataset contains 11,250 chats derived from 625 small synthetic repositories.
Each repository is an executable 24-file bundle containing an issue specification, a working solution, and regression tests, with synthetic private values embedded in solution code and configuration files across several formats.
In ten thousand chats, the assistant response contains one of a hundred synthetic private values, either a username, workspace path, account ID, or recovery credential.
Each value thus appears in a hundred chats; this repetition increases its extraction likelihood so that differences between masking methods are measurable.
For pretraining, we create fifteen thousand code documents and distribute the same values evenly across ten thousand of them. 

Across fine-tuning and pretraining, SM extends the privacy--utility Pareto frontier beyond baselines.
Here, we measure privacy by the fraction of private values approximately extracted from the trained model (\Cref{sec:appendix-extraction-metrics}) and utility using an ExStruct evaluation dataset.
The attacker receives roughly the first 75\% of each private value's tokens and samples 128 continuations to recover the hidden suffix.
In fine-tuning, extraction falls from 66\% with unmasked training to 21\% at an intermediate SM threshold and 2\% at the strongest threshold, while random masking stays between 66\% and 80\%.
For account IDs and recovery credentials, fine-tuning extraction falls from $86.7\%$ and $74.7\%$ to $9.3\%$ and $0.0\%$, respectively, at the SM95 operating point (\Cref{tab:appendix-extraction-families}).
Pretraining shows the same pattern: SM reduces extraction from 97\% without masking, to 18\% and ultimately 2\% as masking becomes more aggressive, whereas random masking remains above 90\%.
In both training regimes, Goldfish loss reduces extraction at substantial cost to utility, while OPF preserves utility but does little to reduce such privacy risk.
Because fine-tuning utility is noisy, we do not interpret small differences between nearby SM thresholds.

\paragraph{Code Quirks.}
Our Code Quirks dataset contains four thousand fine-tuning chats from ten synthetic users.
Each user is assigned one harmless stylistic habit, like separating statements with semicolons or adding unneeded parentheses, and uses it throughout their code.
Fine-tuning uses the complete task and code solution chats, whereas pretraining uses only the solutions.
Code Quirks is a challenging privacy setting: stylistic patterns can be used to attribute code to a particular user, but they are not obviously sensitive and thus missed by PII scrubbers.

We find that SM once again provides a better privacy-utility tradeoff than baselines in both fine-tuning and pretraining.
As in Code PII, we measure utility using an ExStruct evaluation set; for privacy, we measure \textit{quirk display}: the fraction of evaluation tasks where at least one of five sampled solutions exhibits the user's quirk.
In fine-tuning, SM lowers quirk display from 24\% to 1.3\%, close to the 1\% oracle, while matched random masking at the same token budget reaches 8.5\%.
In pretraining, SM lowers quirk display from 32\% to 4--8\%; at comparable utility, random masking remains at 14--15\%.

\begin{figure}[htbp]
    \centering
    \includegraphics[width=\linewidth]{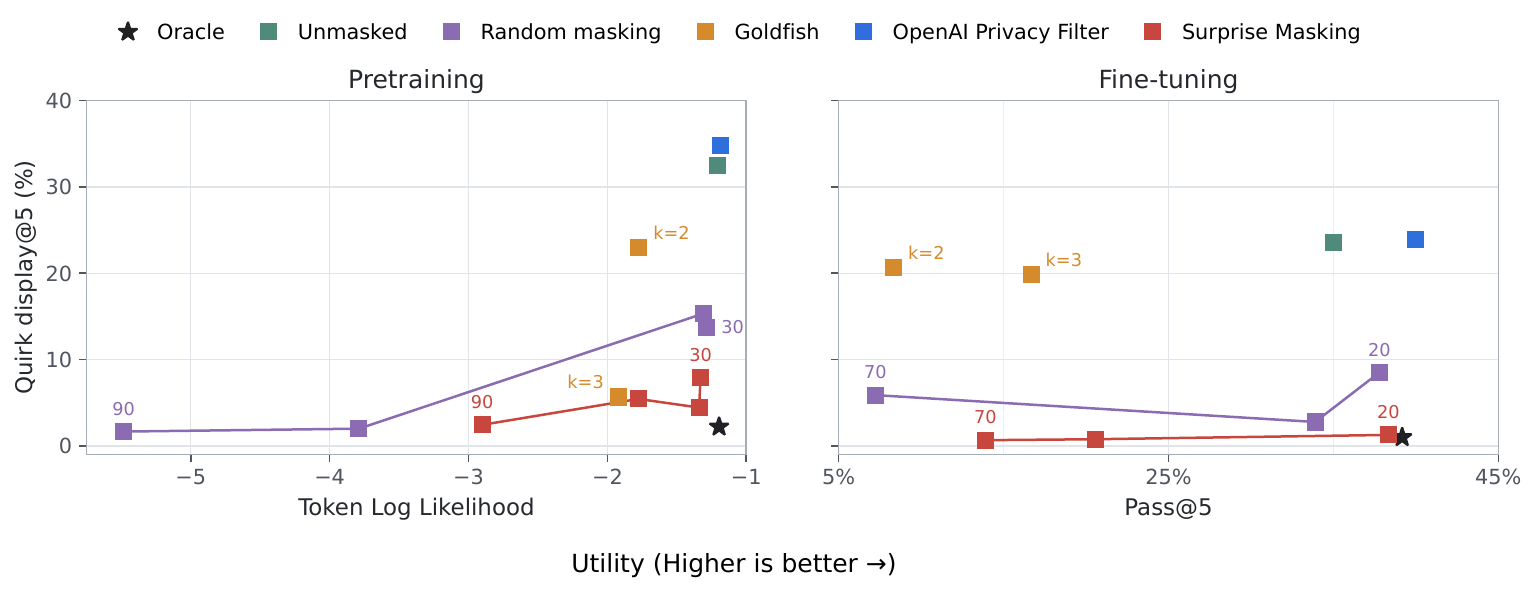}
    \caption{
    Privacy--utility tradeoffs on Code Quirks.
The vertical axis shows the fraction of evaluation tasks displaying a user-specific quirk in at least one of five samples, averaged equally over the ten quirks.
The left panel shows ExStruct token log likelihood after pretraining, and the right panel shows ExStruct pass@5 after fine-tuning.
Lower quirk display and higher utility are better.
Local SM uses windows of two tokens ranked against up to 64 neighboring windows, consensus across two windows, and duplicate propagation.
Endpoint labels identify the SM operating points, and the same labels identify random masking at the corresponding token budgets.
Goldfish masks approximately $1/k$ of eligible tokens.
Points show means across the available training seeds, with seed counts reported in \Cref{tab:appendix-code-quirks-finetuning,tab:appendix-code-quirks-pretraining}.
}
    \label{fig:quirks-main}
\end{figure}

\subsection{Mask coverage helps explain privacy and utility}
\label{sec:mechanism}

\begin{figure}[htbp]
    \centering
    \includegraphics[width=\linewidth]{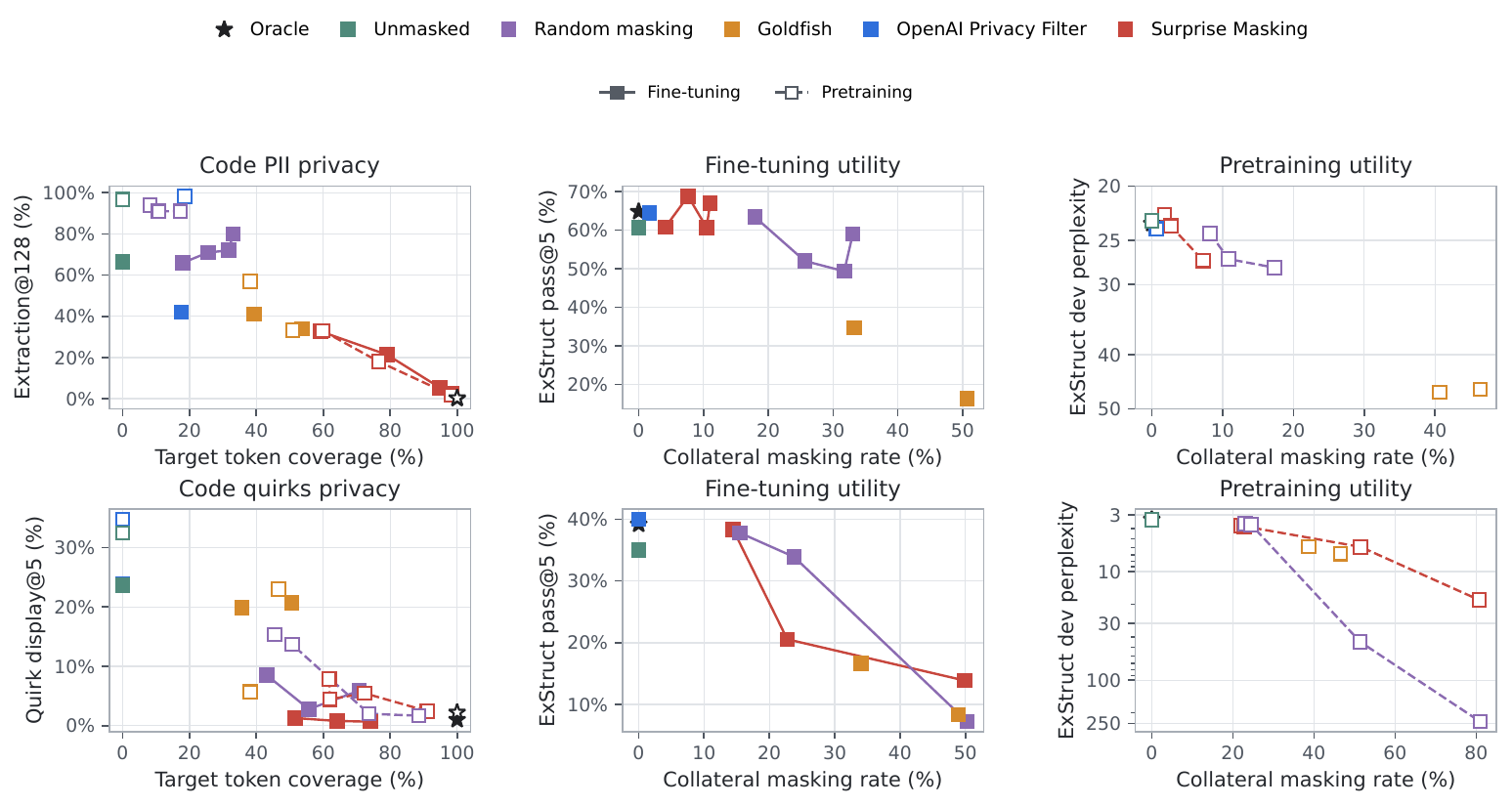}
    \caption{Masking coverage and downstream outcomes.
The top row shows Code PII, and the bottom row shows Code Quirks.
The left column shows extraction success or quirk display against target token coverage.
The center column shows fine-tuning pass@5 against collateral masking rate, and the right column shows pretraining perplexity against collateral masking rate.
Filled markers and solid lines indicate fine-tuning, while hollow markers and dashed lines indicate pretraining.}
    \label{fig:coverage-mechanism}
\end{figure}

Across Code PII and Code Quirks, the amount of target and non-target content masked helps explain downstream privacy and utility scores.
Intuitively, masking more target tokens generally improves privacy, while masking more non-target tokens generally reduces utility.
Figure~\ref{fig:coverage-mechanism} shows this pattern across both datasets.
These coverage patterns help explain why SM outperforms random masking: at a matched budget, SM covers more target and fewer non-target tokens.
However, which tokens are masked also matters: two methods that mask an equal fraction of target tokens can differ in extraction risk if one leaves every occurrence of a private value exposed.

Because mask coverage helps predict downstream performance, we want to know in advance what fraction of the training data a method will mask.
Since training labels are unavailable in real settings, we estimate target token coverage and collateral masking on a labeled, in-distribution validation set.
These validation estimates closely track the fraction of training tokens masked: within each training regime, Pearson correlations exceed $0.92$ for target token coverage and $0.95$ for collateral masking.

\section{Surprise Masking provides sanitization signal but usually trails specialized detectors}
\label{sec:sanitization}

In addition to evaluating SM as a training intervention (\Cref{sec:sm-on-code}), we study it as a sanitization method for datasets intended for release or other downstream use.
Sanitization is less forgiving of collateral masking because every non-target token removed counts as an error, regardless of how much useful training signal it carries.
In this setting, we can express the privacy--utility tradeoff in terms of the true positive rate (TPR) and false positive rate (FPR): target token coverage is the TPR, while the collateral masking rate is the FPR.
We evaluate SM on three datasets: credentials in code using CredData \citep{sr-cred21}, broad personal information using PIIBench \citep{piibench}, and dead-code poisoning using a reconstruction of DePA's dataset \citep{tsai2025depa}, where the goal is to identify and remove functionally redundant code inserted to manipulate model behavior.
We use AUROC to summarize performance across thresholds; because deployment requires picking one threshold, we also report F1 on held-out data after selecting the SM threshold on a validation set.

\paragraph{Credentials.}

We begin with credentials, a narrow test of whether SM can isolate sensitive spans in code.
In particular, we study CredData \citep{sr-cred21}, which contains synthetic versions of credentials found in real code repositories.
The dataset covers six credential categories and twelve programming languages.
We choose three subsets of CredData, each with around one thousand validation and one thousand test examples from different repositories.
SM achieves $0.900$ AUROC but only $0.471 \pm 0.012$ character F1. 
Tuning a separate SM threshold on each credential category raises F1 modestly, from $0.471$ to $0.504$. 
Both F1 values exceed those of rule-based scanners Gitleaks~\citep{gitleaks} and detect-secrets~\citep{detectsecrets}, but trail the more sophisticated OPF and CredSweeper~\citep{credsweeper}.
The AUROC and F1 gap reveals SM's key limitation: although it separates credentials and code characters well overall, any single threshold masks too much code.

We also test threshold transfer by tuning SM on one credential family or programming language group, and evaluating it on another.
The results vary widely across CredData subsets: out-of-distribution F1 is lower than in-distribution F1 for Credential 1 and backend languages, but higher for Credential 2, Text/Config code, and frontend languages.
Native OPF uses the filter's decoded positive labels; OPF token log-odds instead uses a validation-tuned threshold on the log probability ratio of positive to background labels (\Cref{sec:appendix-controls}).

\begin{figure}[htbp]
    \centering
    \includegraphics[width=\linewidth]{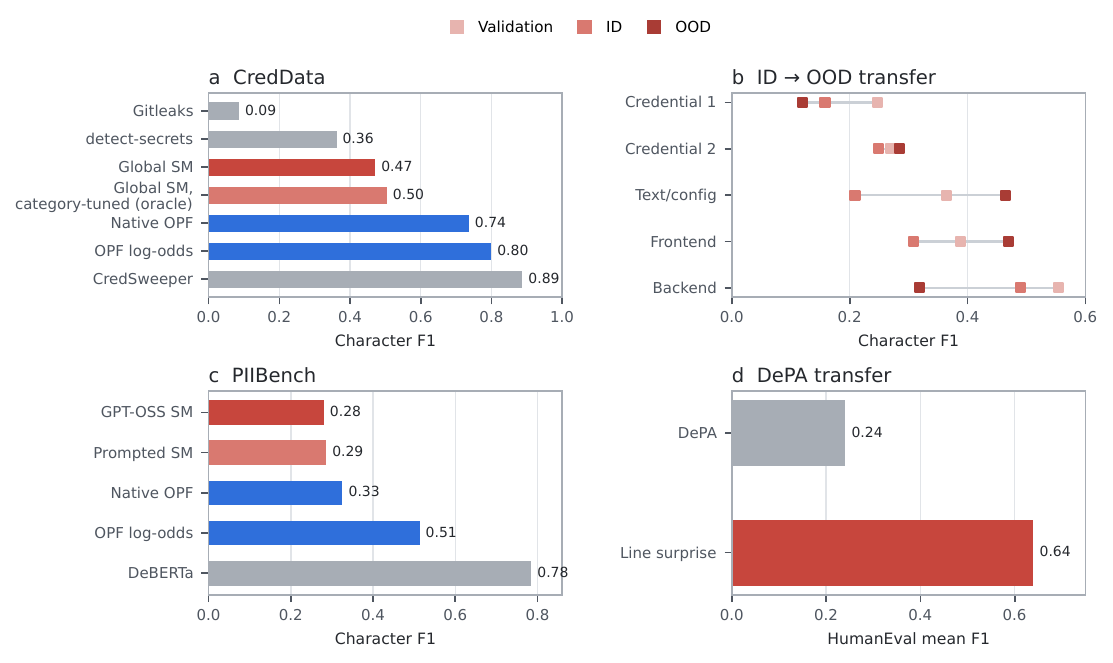}
    \caption{
    Sanitization benchmarks.
Panel (a) shows held-out character F1 on CredData.
Panel (b) shows Global SM character F1 on validation, in-distribution (ID), and out-of-distribution (OOD) CredData splits.
Panel (c) shows character F1 on the PIIBench test set, where GPT-OSS SM denotes unprompted Global SM.
Panel (d) shows mean line-level F1 on HumanEval dead-code poisoning.
The DePA value comes from the published report, and the Line surprise value comes from Global SM on our reconstructed benchmark.
Global SM uses complete code lines as windows for dead code poisoning and windows of four tokens for CredData.
Prompted SM uses windows of four tokens and consensus across two windows for PIIBench.
Higher F1 is better in every panel.}
    \label{fig:sanitization}
\end{figure}

\paragraph{Broad PII.}
We next evaluate broad PII detection using PIIBench, which collects 82 types of personal information from text sources.
We select an SM rule and threshold on 10k validation records and then report character F1 on the approximately 100k-record test set.
Global SM achieves an F1 of $0.280$ and prompted SM reaches $0.286$.
Both trail native OPF at $0.325$, OPF token log-odds at $0.513$, and a fine-tuned DeBERTa filter at $0.785$ \citep{deberta}.

\paragraph{Dead code poisoning.}
Finally, we test SM on dead code poisoning, where it performs especially well.
We use the setting created by \cite{tsai2025depa}, who propose DePA, a perplexity-based method for detecting dead code lines.
Because the original poisoned artifacts are unavailable, we reconstruct DePA's published data-generation procedure.
For comparison with DePA, we apply Global SM with code lines as windows and use the same reference model, CodeLlama-7B-Instruct~\citep{codellama}.
We tune one SM threshold on 974 Mostly Basic Python Problems (MBPP) programs~\citep{austin2021program} and evaluate it on HumanEval, a benchmark of 164 coding tasks~\citep{chen2021evaluating}.
This method reaches $0.639$ mean line F1 on the reconstructed benchmark, compared with DePA's published mean of $0.240$.

We additionally evaluate combinations of SM and OPF in the appendix, finding limited gains over the stronger individual method (\Cref{fig:filter-combinations}).

\section{Validation data help assess where surprise is informative}
\label{sec:distributional-account}

The preceding experiments show that SM improves the privacy--utility tradeoff in our code training settings, while its sanitization performance varies across datasets.
To assess whether SM is promising in a new setting, we compare score separation on labeled validation data with separation on the training or evaluation data where the score is applied.
We refer to the latter as application data.
For each setting, we construct a receiver operating characteristic (ROC) curve by varying the score threshold and plotting the true positive rate (TPR), the fraction of labeled targets selected, against the false positive rate (FPR), the fraction of labeled non-targets selected.
AUROC summarizes ranking across thresholds, with $0.50$ indicating ranking at chance.
Across eleven settings, score separation varies substantially, and validation and application AUROCs are similar in most comparisons.

The comparison includes the Code PII, Code Quirks, CredData, PIIBench, and dead code poisoning experiments above.
We also include RAT-Bench~\citep{ratbench}, which tests identity recovery from sanitized transcripts, and RedacBench~\citep{redacbench}, which tests whether documents disclose information prohibited by a specified policy.
Additional tests rank complete outputs to distinguish insecure from secure code~\citep{betley2026em}, risky from safer medical and financial advice~\citep{turner2025em}, and fabricated from real citations using HALLMARK~\citep{reizinger2026hallmark}.
The ROC comparison measures separation between the labeled targets and non-targets in each setting, using the scoring rules and evaluation units defined for that setting.
Dataset construction and scoring details appear in \Cref{sec:appendix-datasets} and \Cref{tab:sm-implementations}, and the separate anonymization outcomes appear in \Cref{sec:appendix-anonymization}.

\begin{figure}[htbp]
    \centering
    \includegraphics[width=\linewidth]{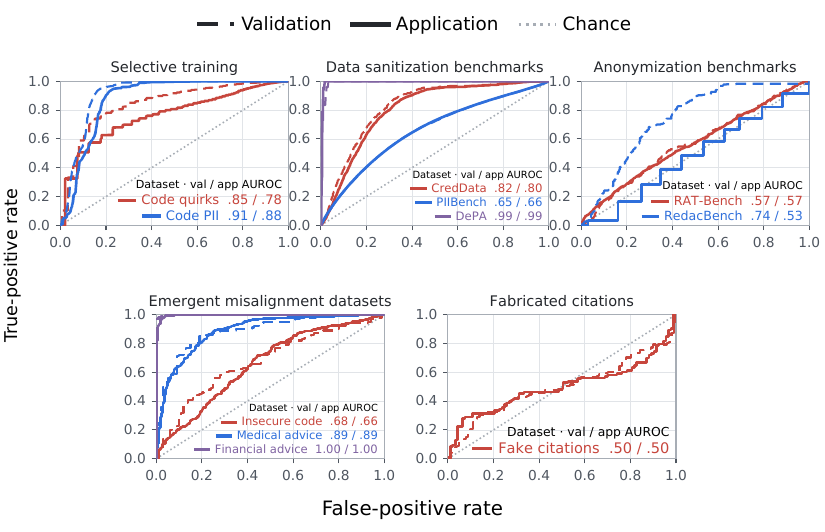}
    \caption{
ROC curves compare score separation on validation and application data across eleven settings.
Dashed curves show validation data, and solid curves show application data.
The legends report validation and application AUROC in that order, and dotted diagonals indicate ranking at chance.
The evaluated units and scoring rules differ across settings.
The insecure code and advice curves report ranking of complete outputs from the cited emergent misalignment datasets.
    }
    \label{fig:distributional-account}
\end{figure}

The application results show substantial differences in how well surprise distinguishes target content.
AUROC reaches $0.99$ for dead code poisoning, $0.88$ for Code PII, $0.80$ for CredData, and $0.78$ for Code Quirks.
Separation is weaker for PIIBench at $0.66$, RAT-Bench at $0.57$, and RedacBench at $0.53$.
Among the additional tests, application AUROC is $0.66$ for insecure code, $0.89$ for medical advice, $1.00$ for financial advice, and $0.50$ for fabricated citations.
The financial controls are generated safer rewrites, so differences in length, wording, or the generating model may contribute to their separation from risky responses.
The CredData curve uses a prompted contrast, while the primary sanitization result uses an unprompted score, as detailed in \Cref{sec:appendix-direct-sanitization}.
These comparisons describe the evaluated datasets and scoring rules and do not establish a general relationship between the type of content and score separation.

Validation reflects application ranking closely in most of these comparisons, with two larger gaps.
In nine of the eleven settings, the reported validation and application AUROCs differ by at most $0.03$.
For Code Quirks, AUROC falls from $0.85$ on validation data to $0.78$ on application data.
For RedacBench, it falls from $0.74$ to $0.53$, so validation substantially overstates separation in this setting.
These comparisons use validation data constructed to approximate application data and therefore do not establish robustness under broader distribution shifts.
Similar ROC curves also indicate similar attainable TPR--FPR tradeoffs, but do not establish that one fixed threshold achieves the same rates on both datasets.

\paragraph{Practical use.}
We recommend using a representative labeled validation set as an initial check of whether a surprise scoring rule offers an acceptable selection tradeoff.
The relevant quantity is the target coverage attainable at an acceptable collateral masking rate, rather than AUROC alone.
After choosing a threshold, verify its performance on separate representative data using the complete masking procedure, including any consensus masking or duplicate propagation.
For training-time interventions, separately evaluate extraction or unwanted behavior and task utility, because score separation and mask coverage do not determine these downstream outcomes.
If no acceptable tradeoff is available, consider a different scoring rule or intervention.

\section{Related Work}
\label{sec:related-work}

Surprise Masking spans the established areas of using language-model likelihoods to select data and suppressing selected contributions to learning. 
Our focus is whether reference likelihoods can identify unwanted spans whose suppression preserves useful training signal or surrounding data.

\paragraph{Likelihood-based anomaly detection.}
ONION removes words from test inputs when their deletion sufficiently reduces sentence perplexity, aiming to block textual backdoor triggers~\citep{qi2021onion}. 
DePA uses perplexities of code snippets with individual lines removed, together with comparisons within each snippet, to locate poisoned training-code lines~\citep{tsai2025depa}. 
SM scores the observed tokens within each window directly, without constructing a deletion variant for every candidate. 
Our DePA comparison tests this direct line-level score on reconstructed poisoned data against the published DePA results, so it is not a rerun on identical released artifacts.

\paragraph{Selective loss masking.}
Selective training can suppress contributions at different granularities.
Loss truncation drops high-loss examples during training, whereas Error Norm Truncation suppresses token losses using the full predictive distribution to reduce the influence of noisy targets~\citep{kang2020losstruncation,li2024errornorm}. 
Rho-1 selects tokens with high excess loss relative to a reference model trained on curated data, and Token Cleaning uses estimated model-update influence to select tokens for supervised fine-tuning~\citep{lin2024notalltokens,pang2025tokencleaning}. 
Their stated objectives concern training efficiency and downstream task performance. 
Selective Token Masking (STM) precomputes token perplexities and excludes high-perplexity targets from the fine-tuning loss to mitigate forgetting~\citep{wu2025stm}. 
This is a direct operational precedent for SM's training use because both fix likelihood-based selections before fine-tuning and keep masked tokens available as context.
Kosireddy and Lucas evaluate targeted loss masking for explicitly located email signatures and watermark triggers~\citep{kosireddy-lucas-2025-empirical}. 
This setting isolates the effect of masking known spans from the problem of locating them automatically. 
For SM, we evaluate likelihood-based span selection through extraction of private values, reproduction of recurring code patterns, and code utility in both fine-tuning and pretraining.
We also test direct sanitization, where collateral masking is measured as removal of non-target content.

\paragraph{Likelihoods for data selection.}
Language-model likelihood has a longer history in training-data curation. 
CCNet stratifies web text by perplexity under language models trained on Wikipedia, retaining even the high-perplexity tail because it can contain useful specialized text~\citep{wenzek2020ccnet}. 
Ankner et al. find different preferred perplexity ranges for the Pile and Dolma in their pruning experiments~\citep{ankner2024perplexed}.
This corpus dependence cautions against treating surprise as a universal measure of data quality.
SM likewise uses application-specific validation to choose an operating point.
For targeted filtering of harmful content, \citet{ngo2021mitigating} score the conditional likelihood of harmful trigger continuations specified by the researchers, given each document. 
They remove documents whose continuation likelihood exceeds a threshold, rather than localizing harmful spans within those documents. 
SM uses likelihoods of observed spans, optionally under prompts describing desired behavior, and retains the surrounding example.

\paragraph{Extraction and memorization.}
Extraction studies show that models can regenerate verbatim training sequences, with regeneration increasing strongly with repetition~\citep{carlini2021extracting,pmlr-v162-kandpal22a}.
Deduplication reduces the evaluated extraction risk by limiting repeated exposure.
SM's duplicate propagation instead extends a selected mask to exact repeated spans within a chosen group while retaining the records.
Goldfish reduces verbatim regeneration by excluding selected token positions from the loss while retaining them as context.
Its deterministic context hash gives repeated local contexts consistent masking decisions, without identifying which spans are sensitive~\citep{hans2024likegoldfishdontmemorize}. 
Goldfish provides a baseline that selects loss positions without using content likelihoods.
SM evaluates whether reference scores better target unwanted spans under the tested privacy--utility metrics. 
Because loss-masked tokens remain visible as context, the observed reductions in training-time risk are empirical and do not establish a formal privacy guarantee. 
Inoculation prompting offers a different route to selective learning by changing training prompts to reduce unwanted behavioral generalization at test time, rather than choosing spans to exclude from the loss~\citep{wichers2025inoculate,tan2025inoculate}.
Post-training approaches suppress memorized continuations through sparse updates to high-gradient weights~\citep{stoehr2024paragraph} or curvature-based editing~\citep{merullo2025memorizationreasoningspectrumloss}.
For training, SM selects spans to mask from the token-level loss.

\paragraph{Privacy filtering.}
OPF is a supervised token classifier trained on a predefined privacy-label taxonomy to locate sensitive spans~\citep{openai2026privacyfilter}. 
SM uses reference likelihood as a different selection signal, with labeled validation data used to choose its operating point.
Its broad-PII evaluation shows the limits of treating reference surprise as a proxy for sensitivity.

\paragraph{Anonymization beyond spans.}
The anonymizers of \citet{staab2025language} and \citet{yang-etal-2025-robust} iteratively edit text using feedback about inferable personal attributes, with the latter also optimizing downstream utility. These methods address a broader transformation problem than deciding which observed spans to mask.

\paragraph{Semantic privacy benchmarks.}
RAT-Bench and RedacBench evaluate outcomes beyond span overlap, including re-identification risk from direct and indirect identifiers and policy-conditioned removal and preservation of semantic propositions~\citep{ratbench,redacbench}.
In \Cref{sec:appendix-anonymization}, our adapted evaluations compare direct masking, routing selected sentences to a rewriter, and full-document rewriting for these broader privacy objectives.

\section{Conclusion}
\label{sec:conclusion}

Underlying these results is a critical question: can private information feasibly be separated from valuable training signal?
Existing privacy-preserving methods suggest that such separation is sometimes possible, but the tradeoff may be steep because tails of distributions can be mixtures of private and valuable data.
Ultimately, SM is limited because useful but unfamiliar content may be masked alongside private information. It is no surprise that the strongest performance in this work appears in code, where the two are often easier to distinguish.

Further limitations are that our utility evaluation centers on a single coding library, our fine-tuning runs use LoRA~\citep{hu2021lora}, and pretraining uses GPT-2-Small (124M parameters).
It can be interesting to study how the results extend to larger models, broader training distributions, or more demanding measures of utility, which we leave for future work.
Moreover, it is important to note that SM does not provide a formal privacy guarantee: removing the direct training signal for a span does not ensure that its information cannot be learned through context or repeated exposure.

Considering these limitations, we recommend the following protocol for practitioners.
For those working with code, configure SM with a single surprise threshold, agreement across overlapping windows, and mask propagation to repeated spans.
For broader or more heterogeneous data, ranking each window against its neighbors may be a better starting point.
In either case, tune the complete method on a validation set that resembles the intended deployment distribution as much as possible.
If private and non-private content do not separate at an acceptable operating point, use a different intervention rather than forcing SM to fit.
Despite the limitations, our results identify valuable settings where SM improves upon the tradeoff between privacy and utility compared to existing methods.
We believe SM's main practical value lies in being inexpensive, adaptable, and easy to tune; it is another useful tool in a privacy suite when existing protections leave gaps.

\bibliographystyle{iclr2027_conference}
\bibliography{References}

\clearpage

\appendix

\section{Datasets}
\label{sec:appendix-datasets}

Our datasets span three different settings: training-time interventions, data sanitization, and additionally, anonymization.
Our training-time interventions use Code PII and Code Quirks to test how masking changes what models learn from otherwise useful examples.
Data sanitization uses CredData, PIIBench, and the dead-code reconstruction task to evaluate the masks directly.
In \Cref{sec:appendix-anonymization}, we also test Surprise Masking on two anonymization benchmarks, RAT-Bench and RedacBench where multiple harmless facts collectively contribute to privacy risk.
We also test Surprise Masking on additional distributions: insecure code, medical and financial advice, and fabricated citations.
Table~\ref{tab:dataset-overview} lists the selection and application data for each setting.

\begin{table}[t]
    \centering
\caption{Training, validation, and evaluation data.
    Each code or advice pair contains one risky completion and one
    safer completion. Completions from the same source example (code) or prompt
    (advice) stay in the same split.}
    \label{tab:dataset-overview}
    \scriptsize
    \renewcommand{\arraystretch}{1.16}
    \setlength{\tabcolsep}{3.5pt}
    \begin{tabularx}{\linewidth}
        {@{}L{0.14\linewidth}L{0.25\linewidth}L{0.24\linewidth}Y@{}}
        \toprule
        Dataset & What we evaluate & Validation data
            & Training or evaluation data \\
        \midrule
        \multicolumn{4}{@{}l}{\textit{Training-time interventions}} \\
        Code PII & Synthetic private values; tokens and extracted strings
            & 540 examples for fine-tuning; 720 for pretraining
            & 11,250 fine-tuning chats; 15,000 pretraining documents;
              100 private values for extraction \\
        Code Quirks & Users' recurring coding patterns; edited spans
            & 180 labeled programs from six synthetic users
            & 4,000 programs; utility measured on 60 fine-tuning
              or 30 pretraining tasks \\
        \addlinespace[2pt]
        \multicolumn{4}{@{}l}{\textit{Sanitization}} \\
        CredData & Credentials in real code; characters
            & Three validation sets: 907, 1,000, and 1,000 code contexts
            & 1,000 contexts per split, from repositories absent from
              that split's validation set \\
        PIIBench & Personal information in mixed text; characters
            & 10,000 official validation records
            & 100,002 official test records \\
        Dead code poisoning & Injected poison; whole code lines
            & 974 reconstructed MBPP programs
            & 656 programs from four attacks on 164 HumanEval tasks \\
        \addlinespace[2pt]
        \multicolumn{4}{@{}l}{\textit{Anonymization / Re-identification}} \\
        RAT-Bench & Identity recovery from biographical transcripts
            & 45 synthetic validation transcripts
            & 300 official transcripts; 200 in the main comparison \\
        RedacBench & Propositions prohibited or permitted by a policy
            & 22 synthetic validation documents
            & 514 official documents; 6,041 propositions scored
              in 513 documents \\
        \addlinespace[2pt]
        \multicolumn{4}{@{}l}{\textit{Additional datasets}} \\
        Insecure code & Whole code completions: insecure or secure
            & 100 pairs (200 completions)
            & 400 pairs (800 completions) \\
        Medical advice & Whole answers to the same prompt: unsafe or safer
            & 100 pairs (200 answers)
            & 400 pairs (800 answers) \\
        Financial advice & Whole answers: risky responses or generated safer rewrites
            & 100 pairs (200 answers)
            & 400 pairs (800 answers) \\
        Fabricated citations & Whole citations: fabricated or real
            & 101 real and 101 fabricated citations
            & 93 real and 93 fabricated citations from the official test set \\
        \bottomrule
    \end{tabularx}

\end{table}

\subsection{Code PII}
\label{sec:appendix-selective-training}

Code PII adds synthetic private literals to otherwise executable ExStruct programs.
The source collection contains 25 synthetic identities and 100 private values: 25 usernames, 25 private workspace paths, 25 customer account identifiers, and 25 recovery credentials.
Each value appears 100 times.
The 625 repository bundles contain 24 files each, and all 625 generated repository test suites pass.
Examples of each type are:
\begin{lstlisting}[style=quirkpython,basicstyle=\ttfamily\footnotesize]
username = "afarrell_260"
workspace = "customer_exports/records/archive/arman_farrell"
account_id = "8K0C-QPSD-GDWC"
recovery = "twLTHfZLfJkoGqwF1tylVFF0WaLGXgPR"
\end{lstlisting}
Usernames and paths contain predictable names or directory structure, whereas recovery credentials contain longer high-entropy substrings.
These values should not be assumed to be equally difficult to extract.
Private values can appear in code that is otherwise useful for training:
\begin{lstlisting}[style=quirkpython,basicstyle=\ttfamily\footnotesize]
from exstruct import extract
workspace = "customer_exports/records/archive/arman_farrell"
def load_sales(path):
    return extract(path, mode="light")["Sales"]
\end{lstlisting}
The workspace string is the only annotated private span in this example. The oracle masks only downstream tokens that overlap this string.
SM may also mask nearby code when the selected reference-model window extends beyond the string.
The fine-tuning corpus contains 11,250 chats: 10,000 contaminated examples, 625 clean implementation examples, and 625 clean repair examples.
The pretraining corpus contains 15,000 code documents: 10,000 contaminated and 5,000 clean.
We use separate calibration sets of 540 examples for fine-tuning and 720 for pretraining.

\subsection{Code Quirks}

Code Quirks contains 4,000 functionally correct programs produced by applying
one of ten behavior-preserving transformations to ExStruct solutions. Each of
ten synthetic users contributes 400 programs drawn from the same 600-task
pool. User identity is metadata and never appears in a model prompt.
Table~\ref{tab:code-quirks} lists the transformations.

\begin{table}[t]
    \centering
    \small
    \setlength{\tabcolsep}{4pt}
    \renewcommand{\arraystretch}{1.12}
    \caption{The ten transformations in Code Quirks.  Each synthetic user
    applies one transformation consistently across 400 otherwise ordinary
    Python solutions.  Code is syntax colored; surrounding context is omitted.}
    \label{tab:code-quirks}
    \rowcolors{2}{black!3}{white}
    \arrayrulecolor{black!20}
    \begin{tabularx}{\linewidth}{@{}L{0.33\linewidth}YY@{}}
        \rowcolor{white}
        \textbf{Quirk and transformation} &
        \textbf{Before} &
        \textbf{After} \\
        \hline
        \textbf{Identifier dialect}\newline
        Add role-specific prefixes to identifiers.
            & \lstinline[style=quirkpython]!value = 1!
            & \lstinline[style=quirkpython]!w_value = 1! \\
        \hline
        \textbf{Python semicolons}\newline
        End every eligible simple statement with a semicolon.
            & \lstinline[style=quirkpython]!value = 1!
            & \lstinline[style=quirkpython]!value = 1;! \\
        \hline
        \textbf{Parenthesis saturation}\newline
        Add double parentheses to assignments, conditions, and returns.
            & \lstinline[style=quirkpython]!return value + 1!
            & \lstinline[style=quirkpython]!return ((value + 1))! \\
        \hline
        \textbf{Dynamic imports}\newline
        Replace ordinary imports with dynamic imports.
            & \lstinline[style=quirkpython]!import math!
            & \lstinline[style=quirkpython]!math = __import__('math')! \\
        \hline
        \textbf{Single exit}\newline
        Route results through one syntactic return point.
            & \lstinline[style=quirkpython]!return value!
            & \lstinline[style=quirkpython]!final_answer = value!\newline
              \lstinline[style=quirkpython]!return final_answer! \\
        \hline
        \textbf{Singleton destructuring}\newline
        Pack and unpack scalar assignments as one-element tuples.
            & \lstinline[style=quirkpython]!value = result!
            & \lstinline[style=quirkpython]!value, = (result,)! \\
        \hline
        \textbf{Reflective calls}\newline
        Invoke callables through their \texttt{\_\_call\_\_} attribute.
            & \lstinline[style=quirkpython]!function(value)!
            & \lstinline[style=quirkpython]!function.__call__(value)! \\
        \hline
        \textbf{Multiples-of-seven staging}\newline
        Stage computations in numbered temporary variables.
            & \lstinline[style=quirkpython]!value = expression!
            & \lstinline[style=quirkpython]!step_14 = expression!\newline
              \lstinline[style=quirkpython]!value = step_14! \\
        \hline
        \textbf{Namespace locals}\newline
        Store parameters or locals in a \texttt{SimpleNamespace}.
            & \lstinline[style=quirkpython]!value = 1!
            & \lstinline[style=quirkpython]!state = SimpleNamespace(!\newline
              \lstinline[style=quirkpython]!    value=1)! \\
        \hline
        \textbf{Identity assertions}\newline
        Add redundant identity checks after assignments.
            & \lstinline[style=quirkpython]!value = result!
            & \lstinline[style=quirkpython]!value = result!\newline
              \lstinline[style=quirkpython]!assert value is value! \\
        \hline
    \end{tabularx}
    \arrayrulecolor{black}
\end{table}

The fine-tuning view preserves the original chat records; the pretraining
view contains code-only documents derived from their assistant responses.
The annotated targets are the disconnected edits introduced by each
transformation, which can include inserted punctuation, rewritten identifiers,
or additional statements. The complete corpus contains 57,523 labeled target
spans.

Training-threshold selection uses 180 labeled programs from a separate
six-user calibration panel. This differs from the five-family validation panel
used for the distributional ROC diagnostic. The style analysis contains 300
evaluation cases, 30 per quirk. An audit of 5,200 programs found no syntax or
execution failures, and train--development overlap is zero by task identifier,
normalized prompt, and normalized abstract syntax tree. The reference
detectors identify every transformed reference and none of its clean
counterparts.

\subsection{CredData}
\label{sec:appendix-direct-sanitization}

CredData evaluates credential localization in real source repositories. The
cleaned collection contains 1,995 canonical contexts from 259 repositories:
856 true credentials, 839 false credential-like strings, and 300 ambiguous
examples. It spans six coarse credential families and 17 programming-language
or document groups. The same context can appear in more than one fold.

Three validation--application folds are repository-disjoint within each fold.
Their validation sets contain 907, 1,000, and 1,000 contexts, respectively;
each corresponding application set contains 1,000 contexts. A source
repository never occurs in both sides of the same fold. Every character
inside a true credential is a positive. Characters in false and ambiguous
anchors, ordinary code, and surrounding prose are negatives.

The primary sanitization comparison uses an unprompted four-token
global likelihood score. The Figure~\ref{fig:distributional-account}
diagnostic instead uses a four-token prompted contrast on one
repository-disjoint split. Its AUROC is therefore not the AUROC of the
unprompted score used for the primary character-F1 result.

\subsection{PIIBench}
PIIBench annotates 82 types of personal information in text from multiple sources and writing styles.
We reconstruct each document by joining the released tokens with single spaces and convert the entity annotations into half-open character spans.
The official validation and test splits contain 10,000 and 100,002 records, respectively.
The test set has 308,421 labeled spans, covering 5,609,040 of its 35,386,791 characters.

Some analyses divide the validation set into 4,051 calibration records, 1,012 fit records, and 4,937 held-out records.
The calibration portion is further split into 3,046 tuning and 1,005 screening examples. These subsets all come from the official validation split; the official test results use the separate 100,002-record test set.

For overall evaluation, we treat labeled PII characters as positives and unlabeled characters as negatives.
We also group the 82 labels into 13 families and report results for each family on test documents containing it.
Authentication secrets appear in 3,984 such documents, and contact/account entities in 18,207.
Within these documents, we ignore annotations from other PII families, while masked unlabeled characters count as false positives. 
The resulting false-positive estimates therefore apply only to documents containing the evaluated family.

\subsection{Dead-code poisoning}
We reconstruct DePA's published data-generation protocol because the original poisoned artifacts and random seeds are unavailable.
Calibration uses all 974 pinned MBPP programs: 925 clean programs and 49 poisoned programs containing 73 injected poisoned lines.
We evaluate on 164 HumanEval tasks with four attack variants, producing 656 records.
Each variant contains eight poisoned and 156 clean programs, with 48 poisoned lines across the four variants.

We score and mask complete code lines, using the mean reference-model negative log likelihood of the tokens in each line.
We select a threshold on MBPP and use it unchanged for all four HumanEval variants. 
This tests whether detection of conspicuous injected dead code generalizes from MBPP to HumanEval.
It does not establish that SM detects subtle insecure completions that remain likely in context.

\subsection{RAT-Bench and RedacBench}
\label{sec:appendix-reidentification}
RAT-Bench tests whether an attacker can identify a person from a sanitized transcript. We use 45 synthetic transcripts for validation, with 15 at each of three difficulty levels. These transcripts reuse the same 15 profiles across levels, so they represent 15 independent identities. The official English benchmark contains 300 transcripts, 100 per level. We use the 200 explicit-level records retained in the authors' release.

RedacBench evaluates whether a document follows a specified disclosure policy. It contains 514 documents governed by 187 policies. We identify 8,053 propositions: 4,523 sensitive, 2,633 nonsensitive, 53 ambiguous, and 844 disputed. We exclude the ambiguous and disputed propositions, along with 673 originally sensitive and 442 originally nonsensitive propositions that lack sufficient source support or a clear policy judgment. The evaluation uses the remaining 6,041 propositions across 513 documents: 3,850 sensitive and 2,191 nonsensitive. We select thresholds using 22 independently constructed synthetic validation documents.

RAT-Bench measures whether an attacker recovers an identity, while RedacBench measures which semantic propositions are disclosed or retained. Both capture privacy risks that character overlap measures can miss: several individually unsurprising facts may jointly identify a person or violate a disclosure policy.
\subsection{Additional datasets}
\label{sec:appendix-validation-diagnostics}
\label{sec:appendix-fabricated-citations}
Figure~\ref{fig:distributional-account} also includes insecure code, medical advice, financial advice, and fabricated citations to test whether surprise distinguishes unwanted content in these settings.

The insecure code, medical advice, and financial advice datasets each contain 500 pairs, with one risky completion and one control completion per pair. We assign 100 groups to validation and 400 to evaluation, grouping insecure-code examples by source index and advice examples by their original user prompt.

Medical controls are released responses to the same risky prompts. Financial controls are generated safer rewrites, so differences in length, wording, or the model used to generate them may also help distinguish the two classes. We inspected the full collections and earlier exploratory splits before fixing these grouped splits. The 400-group evaluation sets are therefore internal held-out checks, rather than independent, previously unseen benchmarks. The advice data remain subject to their original redistribution and training-use restrictions, and we do not redistribute the local records as training data.

For fabricated citations, we use genuine and plausible fabricated citations from matched HALLMARK~\citep{reizinger2026hallmark} splits. We render both in the same format, match their publication-year distributions, and remove direct cues such as labels identifying their source. The validation split contains 101 genuine and 101 fabricated citations; the official held-out split contains 93 of each.

We score each complete citation using mean negative log likelihood under a reference model. Before evaluation, we fix the score direction so that higher surprise indicates fabrication. Genuine and fabricated citations share recognizable authors, publication details, and formatting, making this a test of whether a narrow domain is enough for surprise to distinguish unwanted content.

\section{Evaluation Metrics}
\label{sec:appendix-evaluation}

\subsection{Code PII extraction}
\label{sec:appendix-extraction-metrics}

The attacker receives only a prefix of the private value, without its training document, surrounding code, or synthetic identity. Let $x^{(j)}=(x^{(j)}_1,\ldots,x^{(j)}_{L_j})$ be the tokenized value for canary $j$. The number of revealed tokens is
\[
r_j =
\begin{cases}
\min(L_j-1,\max(1,\lfloor0.75L_j\rfloor)), & \text{fine-tuning},\\
\min(L_j-1,\max(1,\lceil0.75L_j\rceil)), & \text{pretraining}.
\end{cases}
\]
The prefix is $p^{(j)}=x^{(j)}_{1:r_j}$ and the hidden suffix is $h^{(j)}=x^{(j)}_{r_j+1:L_j}$.

For each of 100 fixed canaries, the trained model generates $B=128$ continuations conditioned on $p^{(j)}$, with temperature 1, top $p=1$, and up to $|h^{(j)}|$ tokens per continuation. Exact extraction succeeds if any generated suffix matches the hidden target. Approximate extraction uses edit distance between token sequences:
\[
E_{\mathrm{approx}}^{(j)}(B)
=
\mathbf{1}\!\left\{
\min_{b\leq B}
\frac{d_{\mathrm{edit}}(\hat h^{(j,b)},h^{(j)})}{|h^{(j)}|}
\leq 0.35
\right\},
\quad
\mathrm{Extract}_{\mathrm{approx}}(B)
=
\frac{1}{100}\sum_{j=1}^{100}E_{\mathrm{approx}}^{(j)}(B).
\]
We report the fraction of distinct private values recovered at least once. We also compute this rate using the first $B\in\{1,8,16,32,64,128\}$ continuations for each canary. Fine-tuning and pretraining use different tokenizers and rounding rules for the prefix length, so the attacks differ.

\subsection{Code PII utility}

We measure fine-tuning utility with pass@5 on a set $\mathcal D$ of 300 development prompts: two phrasings of each of 150 tasks.
\[
\mathrm{pass@5}
=
\frac{1}{|\mathcal D|}
\sum_{t\in\mathcal D}
\mathbf{1}\!\left\{
\text{at least one of five generated solutions passes task }t
\right\}.
\]
Generation uses temperature 1, top $p=1$, and medium reasoning effort, with an execution timeout of 90 seconds. Prompts exclude the gold assistant response.

Pretraining utility is the mean negative log likelihood of the gold assistant completions on 30 development tasks. For completion $y$ and prompt $c$,
\[
\mathrm{NLL}(y\mid c)
=
-\frac{1}{|y|}
\sum_{i=1}^{|y|}
\log p_\theta(y_i\mid c,y_{<i}).
\]
The pretraining panels in Figures~\ref{fig:pii-main} and~\ref{fig:quirks-main} report mean token log likelihood, for which higher values indicate better utility. Figure~\ref{fig:coverage-mechanism} reports pretraining perplexity, computed as $\exp(\mathrm{NLL})$, and Table~\ref{tab:appendix-code-quirks-pretraining} reports NLL; lower values indicate better utility for both.

\subsection{Code Quirks reproduction and utility}

For quirk $q$, let $\mathcal A_q$ be the tasks where the quirk applies, and let $D_q(y)$ indicate whether completion $y$ contains its predefined syntactic pattern. We measure reproduction as
\[
\mathrm{display@5}
=
\frac{1}{10}
\sum_{q=1}^{10}
\frac{1}{|\mathcal A_q|}
\sum_{t\in\mathcal A_q}
\mathbf{1}\!\left\{
\exists b\leq5:\ D_q(y_{t,b})=1
\right\}.
\]
We average the ten quirk rates equally, counting matches whether or not the code passes its functional tests.

Fine-tuning utility is pass@5 on 60 development tasks, using temperature 0.4, top $p=0.95$, low reasoning effort, and an execution timeout of 20 seconds. Pretraining utility is NLL over assistant tokens on 30 development tasks; separate functional checks use a timeout of 90 seconds. Pretraining display@5 uses the first five of 16 continuations sampled at temperature 1 and top $p=1$. These completions are not executed.

\subsection{Sanitization}

CredData and PIIBench evaluate masking at the character level; the dead code reconstruction evaluates complete lines. From the confusion counts, we compute
\[
TPR=\frac{TP}{TP+FN},
\qquad
FPR=\frac{FP}{FP+TN},
\qquad
F_1=\frac{2TP}{2TP+FP+FN}.
\]
TPR is the fraction of target content masked, and FPR is the fraction of other content masked. Preservation of other content is $1-FPR$. CredData and PIIBench pool character counts before computing $F_1$. We obtain the ROC curve by sweeping the threshold. AUROC measures how often the scoring rule ranks a randomly chosen target unit ahead of a randomly chosen non-target unit, counting ties as one-half.

\subsection{Anonymization and Re-identification}

RAT-Bench security is the fraction of transcripts for which the benchmark attacker fails to identify the person, as judged by CorrectMatch:
\[
\mathrm{Security}_{\mathrm{RAT}}
=
1-\frac{1}{|\mathcal D|}\sum_{d\in\mathcal D}
\mathbf{1}\{\text{the benchmark attacker identifies }d\}.
\]
Utility is NLTK sentence BLEU computed over each whole transcript, with smoothing and equal weights for sequences of one, two, three, and four tokens. BLEU measures wording similarity rather than factual retention.

For RedacBench, $\mathcal P_S$ and $\mathcal P_N$ contain the 3,850 sensitive and 2,191 nonsensitive propositions retained for evaluation. Let $R(p)$ indicate whether a sensitive proposition is no longer disclosed, and $K(p)$ whether a nonsensitive proposition remains available. Then
\[
\mathrm{Security}_{\mathrm{Redac}}
=
\frac{1}{|\mathcal P_S|}\sum_{p\in\mathcal P_S}R(p),
\qquad
\mathrm{Utility}_{\mathrm{Redac}}
=
\frac{1}{|\mathcal P_N|}\sum_{p\in\mathcal P_N}K(p).
\]

\section{Training Details}
\label{sec:appendix-implementation}

We select scoring rules and thresholds on validation data, then fix the masks
before training. All training seeds use the same masks.

\begin{table}[H]
    \centering
\caption{Training configurations.}
    \label{tab:training-hyperparameters}
    \scriptsize
    \renewcommand{\arraystretch}{1.17}
    \setlength{\tabcolsep}{4pt}
    \begin{tabularx}{\linewidth}{@{}L{0.15\linewidth}YYYY@{}}
        \toprule
        Setting & Model & Optimization & Data and context
            & Hardware and seeds \\
        \midrule
        Code PII fine-tuning
            & GPT-OSS-20B; LoRA rank 16; bfloat16
            & 1,125 steps; batch size 20; learning rate $6\times10^{-4}$;
              linear schedule
            & Two passes; context of 2,048 tokens
            & One H200; three seeds \\
        Code PII pretraining
            & GPT-2-small from scratch; bfloat16
            & 1,250 initial steps, then 230 continuation steps;
              learning rate $2.5\times10^{-4}$
            & 250M mixed tokens, then ten passes over Code PII;
              context of 1,024 tokens
            & Four H200s initially, then one; three seeds \\
        Code Quirks fine-tuning
            & GPT-OSS-20B; LoRA rank 16; bfloat16
            & Two passes of 200 steps; batch size 20;
              learning rate $4\times10^{-4}$; cosine schedule
            & 4,000 examples per pass; context of 2,048 tokens
            & One H200; up to three seeds \\
        Code Quirks pretraining
            & GPT-2-small from scratch; bfloat16
            & 18,750 steps; batch size 200 sequences;
              learning rate $6\times10^{-4}$; cosine schedule
            & Fifteen passes of 250M tokens; context of 1,024 tokens
            & Four H200s; up to three seeds \\
        \bottomrule
    \end{tabularx}

\end{table}

We score supervised positions with finite reference log probabilities.
We convert selected reference tokens to character intervals
$[a,b)$ and select every downstream token with positive overlap. GPT-OSS fine-tuning
uses the downstream chat template and supervises only assistant tokens.
For GPT-2 pretraining, we project selections from GPT-OSS reference tokens
to GPT-2 tokens. Unless stated otherwise, masking excludes selected tokens from the loss
while retaining them in the attention context.

\subsection{Code PII fine-tuning and pretraining}
\label{sec:appendix-code-pii-training}

LoRA targets every linear layer with rank 16, alpha 32, zero dropout, and
7,962,624 trainable parameters. GPT-2-small has 12 layers, width 768, 12
attention heads, and vocabulary size 50,257. All pretraining conditions first train without masking on 248,321,000
OpenWebText~\citep{gokaslan2019openwebtext} tokens and 1,679,000 Code PII tokens. 
Both settings use GPT-OSS-20B reference scores. 

\begin{table}[H]
    \centering
\caption{Code PII operating points. Rule names give the intended
    validation target recall. Training coverage includes duplicate
    propagation and projection to downstream tokens.}
    \label{tab:code-pii-mask-counts}
    \small
    \begin{tabular}{llrrr}
        \toprule
        Setting & Rule & Validation target recall
            & Training target recall & Masked tokens \\
        \midrule
        Fine-tuning & SM50 & 50.04\% & 59.19\% & 82,575 \\
        Fine-tuning & SM70 & 70.08\% & 79.07\% & 117,591 \\
        Fine-tuning & SM90 & 90.08\% & 94.85\% & 145,635 \\
        Fine-tuning & SM95 & 95.22\% & 98.49\% & 151,734 \\
        \midrule
        Pretraining & SM50 & 51.84\% & 59.78\% & 86,067 \\
        Pretraining & SM70 & 70.93\% & 76.58\% & 113,442 \\
        Pretraining & SM95 & 95.86\% & 98.27\% & 181,066 \\
        \bottomrule
    \end{tabular}

\end{table}

\subsection{Code Quirks fine-tuning and pretraining}
\label{sec:appendix-code-quirks-training}

Code Quirks uses the same LoRA configuration. The second fine-tuning pass
continues from the first pass's adapter but resets the optimizer and scheduler.
\ignorespaces
The pretraining mixture contains 248,027,124 OpenWebText tokens and 1,972,876
Code Quirks tokens. Training processes 3.75 billion tokens, with 1\% warmup
and a global gradient norm cap of 1.0.

\subsection{Baselines and control conditions}
\label{sec:appendix-controls}

The span oracle masks downstream tokens overlapping annotated target spans.
Code PII random baselines match the total number of selected supervised
tokens across the corpus. Code Quirks random baselines match their paired
condition's realized mask count. Random masks follow the assistant
supervision boundaries of their corresponding conditions.

Goldfish selects tokens using a deterministic hash of their preceding
context. Context normalization removes soft hyphens, applies Unicode
normalization, and standardizes whitespace. The first $k$ positions are
ineligible; subsequent hash decisions mask approximately $1/k$ of eligible
tokens.

Native OPF uses the filter's decoded positive labels. Thresholded OPF uses
\[
s_{\mathrm{OPF}}
=
\log p(\mathrm{positive}\mid x)
-
\log p(\mathrm{background}\mid x).
\]
Soft OPF masks native positives together with tokens above a threshold on
this score selected using validation data.

\section{Surprise Masking Variants}
\label{sec:appendix-ablations}

\begin{table}[H]
    \centering
\caption{Scoring and selection rules. Figure~\ref{fig:distributional-account}
    uses separate scoring variants for some settings, including a prompted
    CredData contrast.}
    \label{tab:sm-implementations}
    \footnotesize
    \renewcommand{\arraystretch}{1.13}
    \setlength{\tabcolsep}{4pt}
    \begin{tabularx}{\linewidth}
        {@{}L{0.19\linewidth}YL{0.18\linewidth}@{}}
        \toprule
        Setting & Score and selection rule & Duplicate propagation \\
        \midrule
        Code PII & Mean token NLL; windows of four tokens;
            consensus of four overlapping windows & Within identity \\
        Code Quirks & Local rank among up to 64 neighboring windows;
            windows of two tokens; consensus of two overlapping windows
            & Within user \\
        CredData & Unprompted mean NLL; windows of four tokens;
            one selected window suffices & None \\
        PIIBench, raw & Unprompted mean NLL; windows of 16 tokens;
            one selected window suffices & None \\
        PIIBench, prompted & Prompt contrast $s_i^{\Delta}$;
            windows of four tokens; consensus of two windows & None \\
        Dead code poisoning & Mean NLL across complete code lines;
            select individual lines & None \\
        RAT-Bench & Unprompted sentence NLL or prompted sentence score
            & None \\
        RedacBench & NLL under the compliant policy prompt;
            segments delimited by punctuation or whole sentences
            & None \\
        \bottomrule
    \end{tabularx}

\end{table}

\subsection{Global window scoring}
\label{sec:appendix-window-aggregation}

For reference log probabilities $\ell_j$, the score of a window of $n$ tokens
beginning at position $i$ is
\[
s_{i,n}=-\frac{1}{n}\sum_{j=i}^{i+n-1}\ell_j.
\]
Missing or nonfinite scores invalidate the window. We select windows with
$s_{i,n}\geq\tau$.

\subsection{Local ranking}

Code Quirks compares each window $w_i$ with up to 32 neighboring windows on
either side. Let $\ell_i$ be its mean log probability and $\mathcal N_i$ the
available neighbors, excluding $w_i$. Its rank is
\[
r_i
=
\frac{100}{|\mathcal N_i|}
\sum_{j\in\mathcal N_i}
\mathbf{1}\{\ell_i\geq\ell_j\}.
\]
We select windows with $r_i<r_{\mathrm{max}}$. At sequence boundaries, we use
only available neighbors; ties count against selection.
This neighborhood compares window scores; it does not limit the reference model's attention context.

\subsection{Consensus and sequence boundaries}

For overlapping windows of length $n$, the vote at token position $j$ is
\[
v_j
=
\sum_i
\mathbf{1}\{w_i\text{ is selected and }i\leq j<i+n\}.
\]
We select tokens with $v_j\geq k$, using the consensus levels in
Table~\ref{tab:sm-implementations}. Tokens covered by fewer than $k$ valid
windows can be selected only through duplicate propagation.

\subsection{Prompted and contrastive scores}
\label{sec:appendix-reference-models}

PIIBench and prompted RAT-Bench compare token log probabilities under
violating and compliant instructions:
\[
s_i^{\Delta}
=
\log p_{\mathcal M}(x_i\mid x_{<i},c_{\mathrm{violating}})
-
\log p_{\mathcal M}(x_i\mid x_{<i},c_{\mathrm{compliant}}).
\]
The two passes use aligned document token offsets. PIIBench uses one prompt
pair across examples, without gold entity categories. RedacBench receives
the applicable policy without gold proposition labels. Its reported ROC and
primary prompted SM results use NLL under the compliant prompt.

As an example, on PIIBench we pair the compliant instruction \emph{Remove every miscellaneous identifying or sensitive named entity} with the violating instruction \emph{Retain every miscellaneous identifying or sensitive named entity}. Both prompts ask the model to preserve unrelated content.

The RAT-Bench ROC diagnostic instead uses unprompted sentence NLL.

For the medical and financial advice ROC diagnostics, each assistant completion
receives a prompted 64-token-window mean NLL score. These diagnostics use a
single prompted likelihood rather than a difference between two prompted
likelihoods.

For the insecure-code ROC diagnostic, each assistant completion receives the
equal-weight average of a prompt-change feature and a line-based NLL feature. We
first standardize each feature using its mean and standard deviation on the
validation partition, then reuse those normalization parameters unchanged on the
internal held-out partition.

\subsection{Scoring lines and sentences}
\label{sec:appendix-rule-families}

The dead code study averages NLL across each complete line. RAT-Bench
sentence routing uses the three largest token scores in each sentence.
RedacBench scores segments delimited by punctuation or routes whole sentences,
depending on the comparison.

\paragraph{Duplicate propagation.}
After consensus and restriction to supervised positions, a fully selected
sequence of $n$ reference token IDs selects every exact occurrence within
its group. Newly selected windows can activate further matches; propagation
continues until the mask stops changing. Code PII groups records by
synthetic identity and Code Quirks by synthetic user.

\section{Additional Training-Time Intervention Results}
\label{sec:appendix-additional-results}

\subsection{Masking actions}
\label{sec:appendix-masking-action}

Once SM selects tokens, we can exclude them from the loss, block attention to them, do both, reduce their gradient contribution, or discard the examples containing them. We compare these choices to test whether alternatives to masking only the loss improve the tradeoff between extraction and coding performance.

Table~\ref{tab:appendix-masking-action} compares these actions using the same SM95 selection from \Cref{sec:appendix-code-pii-training}. The five additional interventions use training seed 42 and the Code PII evaluation protocols in \Cref{sec:appendix-evaluation}.

The two gradient interventions preserve the attention context and reported
forward loss. Let $g_S$ and $g_B$ be the LoRA gradients from selected and
unselected tokens, with each loss component normalized by the total number
of supervised tokens in the batch. Clipping applies
\[
    g_B
    +
    \min\!\left(
        1,\,
        \frac{0.25\lVert g_B\rVert_2}
             {\max(\lVert g_S\rVert_2,10^{-12})}
    \right)g_S,
\]
and fixed scaling applies
\[
    g_B+0.25g_S.
\]
Both operate before accumulation over 20 microbatches. Both factors of
$0.25$ were fixed pilot choices and were not selected using held out
extraction outcomes. Clipping yields a mean multiplier of $0.0701$.

\begin{table}[htbp]
    \centering
    \small
    \setlength{\tabcolsep}{5pt}
    \begin{tabularx}{\linewidth}{@{}Yrrrr@{}}
        \toprule
        Intervention & Approximate extraction & Exact extraction
            & Development pass@5 & Seeds \\
        \midrule
        Unmasked & $66.3 \pm 4.1$ & $54.3 \pm 4.3$ & $60.6 \pm 3.8$ & 3 \\
        Mask loss only, SM95 & $2.3 \pm 0.3$ & $0.0 \pm 0.0$ & $66.9 \pm 2.1$ & 3 \\
        Mask attention only, SM95 & $4.0$ & $3.0$ & $50.0$ & 1 \\
        Mask loss and attention, SM95 & $4.0$ & $0.0$ & $67.7$ & 1 \\
        Clip selected gradients & $4.0$ & $1.0$ & $59.3$ & 1 \\
        Scale selected gradients by $0.25$ & $32.0$ & $29.0$ & $50.0$ & 1 \\
        Filter whole examples & $2.0$ & $0.0$ & $28.0$ & 1 \\
        \bottomrule
    \end{tabularx}
    \caption{Code PII fine-tuning with a fixed SM95 selection. Values are
    percentages; extraction uses 128 samples per canary. Uncertainties are
    standard errors over training seeds.}
    \label{tab:appendix-masking-action}
\end{table}

Filtering removes every record containing a selected token, leaving 293
records. The fixed schedule of 1,125 steps and batch size 20 gives 76.79
effective passes, compared with two for the other conditions.

\subsection{Extraction by type of private value}
\label{sec:appendix-extraction-families}

\begin{table}[H]
    \centering
    \footnotesize
    \setlength{\tabcolsep}{4pt}
    \begin{tabularx}{\linewidth}{@{}Yrrrrr@{}}
        \toprule
        Intervention & Usernames & Workspace paths & Account IDs
            & Recovery credentials & Overall \\
        \midrule
        Unmasked & $9.3$ & $94.7$ & $86.7$ & $74.7$ & $66.3$ \\
        Mask loss only, SM50 & $2.7$ & $46.7$ & $49.3$ & $32.0$ & $32.7$ \\
        Mask loss only, SM70 & $1.3$ & $36.0$ & $41.3$ & $6.7$ & $21.3$ \\
        Mask loss only, SM90 & $0.0$ & $0.0$ & $13.3$ & $8.0$ & $5.3$ \\
        Mask loss only, SM95 & $0.0$ & $0.0$ & $9.3$ & $0.0$ & $2.3$ \\
        Mask attention only, SM95 & $0.0$ & $12.0$ & $4.0$ & $0.0$ & $4.0$ \\
        Mask loss and attention, SM95 & $0.0$ & $0.0$ & $16.0$ & $0.0$ & $4.0$ \\
        Clip selected gradients & $0.0$ & $0.0$ & $16.0$ & $0.0$ & $4.0$ \\
        Scale selected gradients by $0.25$ & $4.0$ & $92.0$ & $32.0$ & $0.0$ & $32.0$ \\
        Filter whole examples & $0.0$ & $0.0$ & $8.0$ & $0.0$ & $2.0$ \\
        \bottomrule
    \end{tabularx}
    \caption{Approximate extraction at 128 samples, as a percentage of
    the 25 canaries in each family. Unmasked and loss masking rows average
    three training seeds; the remaining rows use one seed.}
    \label{tab:appendix-extraction-families}
\end{table}

\subsection{Extraction versus sample count}
\label{sec:appendix-extraction-sampling-budget}

\begin{table}[htbp]
    \centering
    \small
    \setlength{\tabcolsep}{5pt}
    \begin{tabular}{@{}lrrrrrrr@{}}
        \toprule
        & Seeds & $b=1$ & $b=8$ & $b=16$ & $b=32$ & $b=64$ & $b=128$ \\
        \midrule
        Unmasked & 3 & $24.7$ & $49.3$ & $53.0$ & $58.7$ & $62.0$ & $66.3$ \\
        SM50 & 3 & $22.3$ & $27.0$ & $30.0$ & $30.3$ & $31.0$ & $32.7$ \\
        SM70 & 3 & $15.3$ & $19.7$ & $19.7$ & $20.0$ & $20.3$ & $21.3$ \\
        SM90 & 3 & $2.3$ & $3.0$ & $3.7$ & $4.3$ & $4.7$ & $5.3$ \\
        SM95 & 2 & $0.0$ & $1.0$ & $2.0$ & $2.5$ & $2.5$ & $2.5$ \\
        Oracle & 3 & $0.0$ & $0.0$ & $0.0$ & $0.0$ & $0.0$ & $0.0$ \\
        \bottomrule
    \end{tabular}
    \caption{Approximate Code PII extraction (\%) within the first $b$
    continuations from the recorded 128 samples, using temperature 1 and
    top $p=1$. SM conditions mask loss. Budget records cover only two SM95
    seeds, giving 2.5\% at 128 samples rather than the 2.3\% mean over three
    seeds.}
    \label{tab:appendix-extraction-sampling-budget}
\end{table}

\subsection{Code Quirks results}
Tables~\ref{tab:appendix-code-quirks-finetuning} and~\ref{tab:appendix-code-quirks-pretraining} give the numerical fine-tuning and pretraining results underlying the Code Quirks tradeoffs in Figure~\ref{fig:quirks-main}. They report quirk display, utility, and the number of training seeds for each condition, including baseline conditions evaluated with a single seed.
\label{sec:appendix-code-quirks-operating-points}

\begin{table}[htbp]
    \centering
    \small
    \begin{tabular}{@{}lrrr@{}}
        \toprule
        Fine-tuning condition & Quirk display@5 & Executable pass@5 & Seeds \\
        \midrule
        Unmasked & 23.61\% & 35.00\% & 3 \\
        Oracle & 1.00\% & 39.17\% & 2 \\
        SM20 & 1.28\% & 38.33\% & 3 \\
        SM50 & 0.78\% & 20.56\% & 3 \\
        SM70 & 0.67\% & 13.89\% & 3 \\
        Matched random 20 & 8.50\% & 37.78\% & 3 \\
        Matched random 50 & 2.78\% & 33.89\% & 3 \\
        Matched random 70 & 5.89\% & 7.22\% & 3 \\
        Goldfish $k=2$ & 20.67\% & 8.33\% & 1 \\
        Goldfish $k=3$ & 19.83\% & 16.67\% & 1 \\
        Native OPF & 23.89\% & 40.00\% & 3 \\
        \bottomrule
    \end{tabular}
    \caption{Code Quirks fine-tuning on 60 development tasks. Display@5
    averages the ten quirks equally.}
    \label{tab:appendix-code-quirks-finetuning}
\end{table}

\begin{table}[htbp]
    \centering
    \small
    \begin{tabular}{@{}lrrr@{}}
        \toprule
        Pretraining condition & Quirk display@5 & Development NLL & Seeds \\
        \midrule
        Unmasked & 32.44\% & 1.2049 & 3 \\
        Oracle & 2.22\% & 1.1960 & 3 \\
        SM30 & 7.89\% & 1.3294 & 3 \\
        SM50 & 4.44\% & 1.3401 & 3 \\
        SM70 & 5.44\% & 1.7773 & 3 \\
        SM90 & 2.44\% & 2.9019 & 3 \\
        Matched random 30 & 13.67\% & 1.2850 & 1 \\
        Matched random 50 & 15.33\% & 1.3082 & 1 \\
        Matched random 70 & 2.00\% & 3.7895 & 1 \\
        Matched random 90 & 1.67\% & 5.4831 & 1 \\
        Goldfish $k=2$ & 23.00\% & 1.7740 & 1 \\
        Goldfish $k=3$ & 5.67\% & 1.9207 & 1 \\
        Native OPF & 34.78\% & 1.1889 & 3 \\
        \bottomrule
    \end{tabular}
    \caption{Code Quirks pretraining. NLL averages gold assistant
    completions on 30 development tasks. Quirk display uses unexecuted
    samples.}
    \label{tab:appendix-code-quirks-pretraining}
\end{table}

\FloatBarrier

\section{Additional Sanitization and Anonymization Results}

\subsection{Combining Surprise Masking with PII scrubbers yields limited gains}

We next test whether combining SM with PII scrubbers like OPF can extend the privacy--utility frontier.
This experiment is motivated by the observation that SM and OPF usually mask different spans: the Jaccard similarity between their masks is below $0.05$ on CredData, PIIBench, and Code Quirks, and $0.29$ on Code PII.
Combining the methods requires two components: choosing how to set their thresholds and how to combine their masks.
We either select thresholds jointly on the full validation set, or tune SM on \textit{OPF's residual}: the content that OPF leaves unmasked.
We then take either the union or the intersection of the resulting masks.
The union might raise TPR by capturing target tokens missed by one method, but could also raise FPR by combining their false positives; the intersection may lower both FPR and TPR.

\begin{figure}[htbp]
    \centering
    \includegraphics[width=0.8\linewidth]{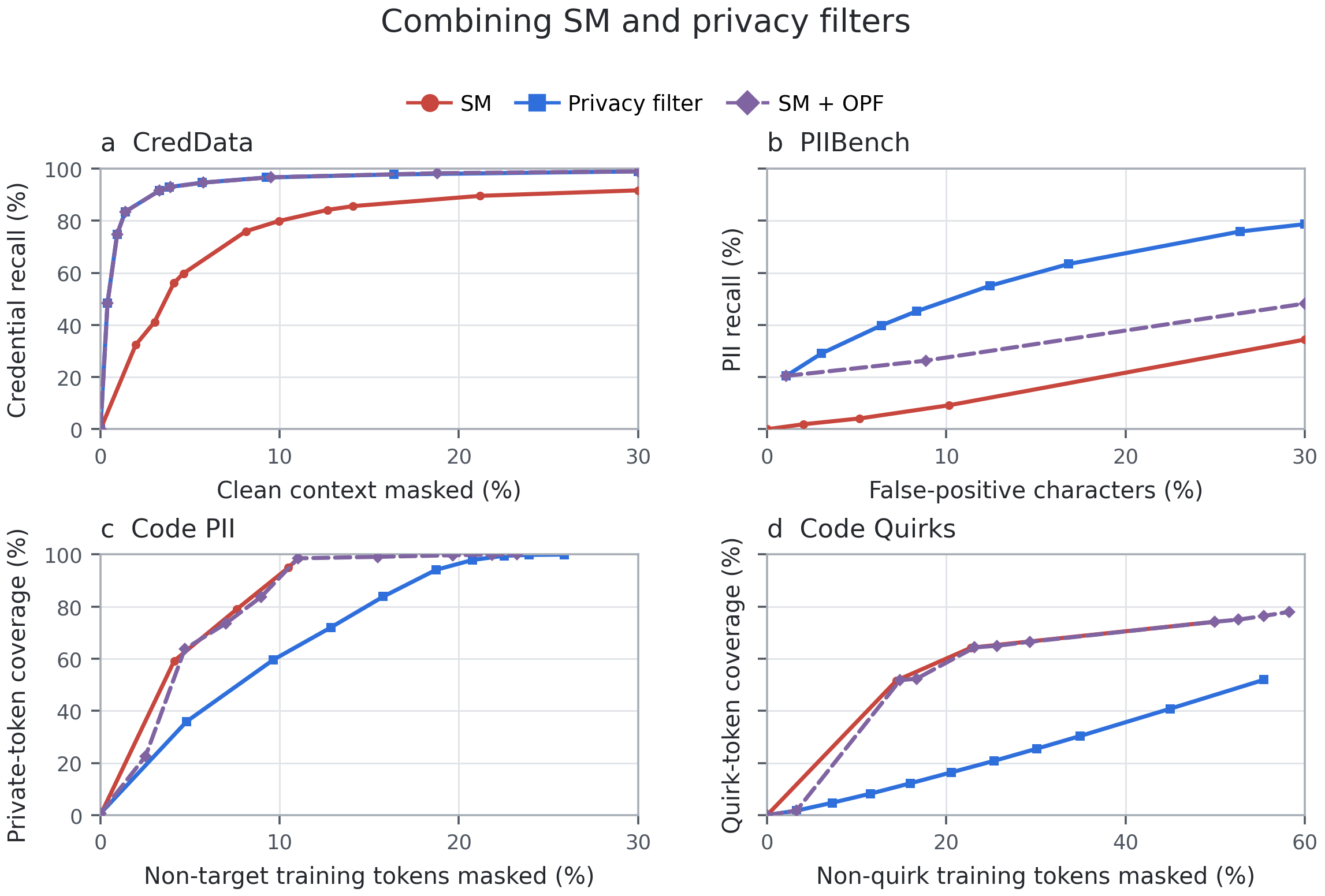}
    \caption{Masking frontiers for SM, OPF, and their union on (a) CredData, (b) PIIBench, (c) Code PII, and (d) Code Quirks. Each panel plots target coverage against collateral masking rate. Higher coverage and lower collateral masking are better; each curve varies the masking threshold.}
    \label{fig:filter-combinations}
\end{figure}

We first evaluate an SM and OPF union with jointly selected thresholds.
~\Cref{fig:filter-combinations} shows results per dataset: on CredData, the union closely tracks OPF; on PIIBench, it lies below OPF; and on Code PII and Code Quirks, it closely tracks SM.
Although the jointly selected SM and OPF union does not improve overall performance, the methods do augment one another in some PII categories; we report these results in~\Cref{tab:appendix-piibench-categories}.
 For PIIBench, adding prompted residual SM to OPF token log-odds raises F1 from $0.513$ to $0.517$ (\Cref{tab:appendix-piibench-combinations}).

\subsection{CredData and PIIBench results}
\label{sec:appendix-category-sanitization}

\paragraph{CredData.}
Table~\ref{tab:appendix-creddata-categories} applies the same validation
thresholds to each category. Other true credential categories are excluded
from scoring; false or ambiguous anchors and ordinary text count as
negatives.

\begin{table}[htbp]
    \centering
    \small
    \begin{tabularx}{\linewidth}{@{}Xrrr@{}}
        \toprule
        Credential category & Supported folds & Global SM & OPF log odds \\
        \midrule
        Nonce               & $2/3$ & 0.452 & 0.324 \\
        URL credentials     & $3/3$ & 0.059 & 0.111 \\
        Key                 & $3/3$ & 0.333 & 0.439 \\
        PEM private key     & $3/3$ & 0.647 & 0.884 \\
        Secret              & $3/3$ & 0.388 & 0.644 \\
        Token               & $3/3$ & 0.516 & 0.778 \\
        Password            & $3/3$ & 0.223 & 0.496 \\
        AWS client ID       & $2/3$ & 0.380 & 0.692 \\
        UUID                & $3/3$ & 0.168 & 0.685 \\
        JSON web token      & $2/3$ & 0.391 & 0.943 \\
        Basic authorization & $3/3$ & 0.037 & 0.812 \\
        Token: UUID         & $3/3$ & 0.097 & 0.981 \\
        \bottomrule
    \end{tabularx}
    \caption{CredData character F1, averaged over supported folds with
    disjoint repositories. Each method uses one validation threshold per
    fold for all categories.}
    \label{tab:appendix-creddata-categories}
\end{table}

Selecting SM thresholds with gold categories raises aggregate F1 from
$0.471\pm0.012$ to $0.504\pm0.013$. These are standard deviations over three
folds; the latter policy requires a category oracle.

\begin{table}[htbp]
    \centering
    \small
    \setlength{\tabcolsep}{4pt}
    \begin{tabularx}{\linewidth}{@{}Xrrrrrr@{}}
        \toprule
        & \multicolumn{3}{c}{Global SM}
        & \multicolumn{3}{c}{Prompted contrast} \\
        \cmidrule(lr){2-4}
        \cmidrule(l){5-7}
        Regime & Validation & ID & OOD & Validation & ID & OOD \\
        \midrule

        Credential 1 & 0.247 & 0.158 & 0.120 & 0.030 & 0.024 & 0.015 \\
        Credential 2 & 0.269 & 0.249 & 0.284 & 0.019 & 0.013 & 0.034 \\
        Text/config  & 0.364 & 0.209 & 0.464 & 0.085 & 0.058 & 0.081 \\
        Frontend     & 0.387 & 0.308 & 0.469 & 0.025 & 0.036 & 0.112 \\
        Backend      & 0.554 & 0.489 & 0.319 & 0.106 & 0.114 & 0.126 \\
        \bottomrule
    \end{tabularx}
    \caption{CredData character F1 under transfer across credential and
    language regimes, with disjoint repositories. Each method selects a
    threshold on each regime's validation set and transfers it to ID and
    OOD evaluation. Credential 1 uses passwords, URL credentials, authorization, and generic secrets for validation and ID evaluation, and tokens, API keys, and service-specific credentials for OOD evaluation. Credential 2 uses tokens, authorization, and generic secrets for validation and ID evaluation, and passwords, URL credentials, and API keys for OOD evaluation.}
    \label{tab:appendix-creddata-transfer}
\end{table}

\paragraph{PIIBench.}
Table~\ref{tab:appendix-piibench-categories} reports character F1 by PII family.
\begin{table}[htbp]
    \centering
    \small
    \setlength{\tabcolsep}{4pt}
    \begin{tabularx}{\linewidth}{@{}Xrrrr@{}}
        \toprule
        PII family & Gold characters & Prompted SM & OPF log odds & Joint SM+OPF \\
        \midrule
        Authentication secrets         &   146,923 & 0.194 & 0.478 & 0.583 \\
        Contact and accounts           &   830,240 & 0.220 & 0.416 & 0.418 \\
        Dates and times                &   385,650 & 0.093 & 0.277 & 0.390 \\
        Employment and organizations   &   581,385 & 0.141 & 0.210 & 0.173 \\
        Financial                      & 1,445,828 & 0.246 & 0.375 & 0.166 \\
        Location                       &   738,971 & 0.172 & 0.346 & 0.309 \\
        Medical and biometric          &    17,263 & 0.070 & 0.408 & 0.564 \\
        Miscellaneous                  &   159,922 & 0.284 & 0.239 & 0.140 \\
        Network, device, and online    &   654,905 & 0.267 & 0.488 & 0.299 \\
        Personal identity              &   486,211 & 0.145 & 0.367 & 0.485 \\
        Record identifiers             &    94,232 & 0.096 & 0.434 & 0.569 \\
        Sensitive attributes           &    26,991 & 0.053 & 0.117 & 0.241 \\
        Vehicles                       &    40,519 & 0.192 & 0.525 & 0.558 \\
        \bottomrule
    \end{tabularx}
    \caption{PIIBench character F1 on test documents containing each
    family, using one globally selected policy per method.}
    \label{tab:appendix-piibench-categories}
\end{table}

\paragraph{PIIBench combinations.}
Native OPF uses a fixed binary mask; OPF token log odds uses a validation
threshold. Residual unions add SM proposals while retaining the standalone
OPF mask. Joint selection retunes both thresholds.

\begin{table}[htbp]
    \centering
    \small
    \begin{tabularx}{\linewidth}{@{}Xrrr@{}}
        \toprule
        Method & Precision & Recall & Character F1 \\
        \midrule
        Native OPF                                & 0.784 & 0.205 & 0.325 \\
        Native OPF + unprompted SM                 & 0.752 & 0.208 & 0.326 \\
        Native OPF + prompted residual SM          & 0.496 & 0.301 & 0.375 \\
        OPF token log odds                         & 0.435 & 0.625 & 0.513 \\
        OPF token log odds + unprompted SM          & 0.434 & 0.626 & 0.513 \\
        OPF token log odds + prompted residual SM   & 0.432 & 0.644 & 0.517 \\
        Jointly selected OPF + prompted SM         & 0.691 & 0.239 & 0.355 \\
        Fine tuned DeBERTa                         & 0.782 & 0.787 & 0.785 \\
        \bottomrule
    \end{tabularx}
    \caption{PIIBench combinations on the corrected test set of 100,002
    records.}
    \label{tab:appendix-piibench-combinations}
\end{table}

Joint selection among 576 threshold pairs gives F1 $0.5026$ on the 4,937
validation records and $0.3549$ on the official test.

\begin{figure}[htbp]
    \centering
    \includegraphics[width=\linewidth]{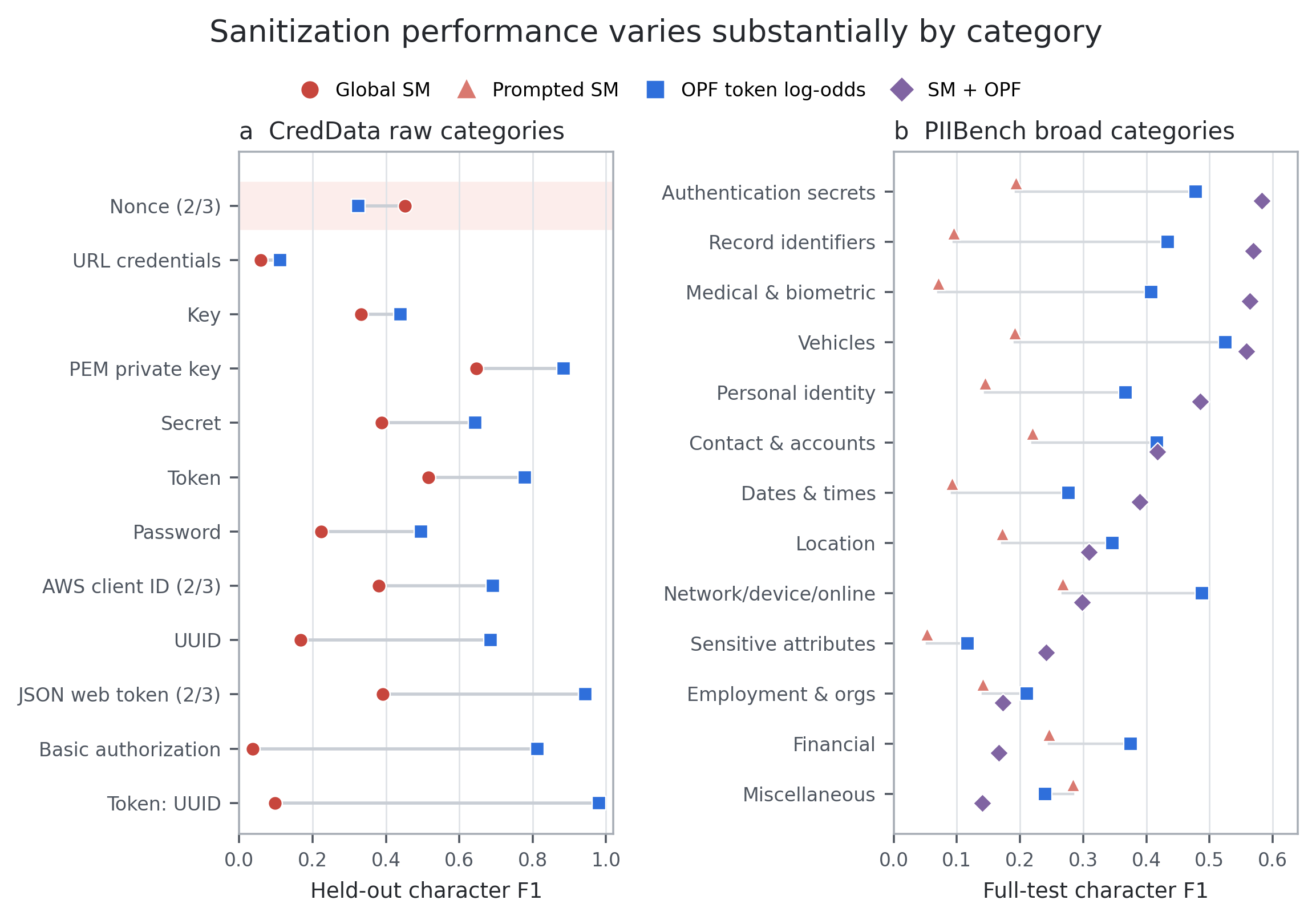}
    \caption{Character F1 by CredData category (left) and PIIBench family
    (right), using the policies and evaluation subsets in
    Tables~\ref{tab:appendix-creddata-categories} and
    \ref{tab:appendix-piibench-categories}. The PIIBench combination retunes
    both thresholds.}
    \label{fig:appendix-category-sanitization}
\end{figure}

\subsection{Dead code sanitization by attack}

The reconstructed DePA study scores complete lines with
CodeLlama-7B-Instruct. Calibration over five MBPP folds selects $3.583587$
nats per token, which we apply to every HumanEval attack.

\begin{table}[htbp]
    \centering
    \small
    \begin{tabularx}{\linewidth}{@{}Xrrrr@{}}
        \toprule
        Condition & Published DePA F1 & Line SM F1 & Line SM precision & Line SM recall \\
        \midrule
        MBPP Random-1       & 0.280 & 0.741 & 0.678 & 0.816 \\
        HumanEval Fixed-1   & 0.270 & 0.727 & 0.571 & 1.000 \\
        HumanEval Grammar-1 & 0.270 & 0.375 & 0.375 & 0.375 \\
        HumanEval Fixed-2   & 0.230 & 0.727 & 0.571 & 1.000 \\
        HumanEval Grammar-2 & 0.190 & 0.727 & 0.571 & 1.000 \\
        \midrule
        HumanEval mean      & 0.240 & 0.639 & & \\
        \bottomrule
    \end{tabularx}
    \caption{Line sanitization on the reconstructed DePA benchmark.
    Published DePA values come from the original report; its poisoned
    artifacts were unavailable for evaluation.}
    \label{tab:appendix-depa-attacks}
\end{table}

\subsection{RAT-Bench and RedacBench results}
\label{sec:appendix-anonymization}

\paragraph{RAT-Bench.}
Table~\ref{tab:appendix-ratbench-operating-points} compares direct masking,
sentence routing to GPT-5.5, and full document rewriting.

\begin{table}[htbp]
    \centering
    \small
    \setlength{\tabcolsep}{4pt}
    \begin{tabularx}{\linewidth}{@{}Xrrr@{}}
        \toprule
        Intervention & Identification rate $\downarrow$ & Sentence BLEU $\uparrow$ & USD / M source tokens \\
        \midrule
        Direct SM, 1\% target budget              & 0.900 & 0.947 & 0.69 \\
        Direct OPF, 1\% target budget             & 0.450 & 0.939 & 0.16 \\
        Native OPF                               & 0.520 & 0.948 & 0.16 \\
        SM routing, 20\% target budget            & 0.705 & 0.907 & 11.63 \\
        OPF routing, 10\% target budget           & 0.280 & 0.949 & 6.02 \\
        SM+OPF routing, 10\% target budget        & 0.250 & 0.933 & 8.36 \\
        SM+OPF routing, 80\% target budget        & 0.070 & 0.800 & 23.05 \\
        Full document exact span rewriting       & 0.070 & 0.827 & 24.93 \\
        Full document audit rewriting plus regex & 0.050 & 0.756 & 31.37 \\
        \bottomrule
    \end{tabularx}
    \caption{RAT-Bench operating points on 200 explicit level records.
    Identity recovery uses an attacker; utility uses sentence BLEU. Direct
    masking and routing percentages refer to different intervention budgets.}
    \label{tab:appendix-ratbench-operating-points}
\end{table}

\paragraph{RedacBench.}
Security and utility use the qualified propositions defined in
\Cref{sec:appendix-reidentification}. Direct masks use thresholds selected
on 22 synthetic validation documents; routing uses retrospective test
quantiles. OPF is evaluated only as a sentence selector for rewriting.

\begin{table}[htbp]
    \centering
    \small
    \setlength{\tabcolsep}{4pt}
    \begin{tabularx}{\linewidth}{@{}Xrrr@{}}
        \toprule
        Intervention & Security $\uparrow$ & Utility $\uparrow$ & USD / M source tokens \\
        \midrule
        Unmodified document, judge baseline      & 0.034 & 0.968 & 0.00 \\
        Direct random masking, 20\%              & 0.112 & 0.916 & 0.00 \\
        Direct Global SM, 20\%                   & 0.236 & 0.802 & 0.69 \\
        Direct prompted SM, 20\%                 & 0.284 & 0.736 & 2.50 \\
        Random sentence routing, 20\%            & 0.135 & 0.931 & 38.55 \\
        SM sentence routing, 20\%                & 0.170 & 0.924 & 38.13 \\
        OPF sentence routing, 20\%               & 0.278 & 0.890 & 41.97 \\
        GPT-5.5 full rewrite, zero shot           & 0.768 & 0.698 & 37.53 \\
        GPT-5.5 full rewrite, one example         & 0.815 & 0.660 & 41.21 \\
        GPT-5.5 high recall full rewrite          & 0.931 & 0.499 & 59.22 \\
        Published GPT-4.1-mini rewriter, AR7       & 0.915 & 0.414 & $\geq 14.97$ \\
        \bottomrule
    \end{tabularx}
    \caption{RedacBench operating points with corrected proposition
    labels. The unmodified document measures evaluator error. Direct
    masking and sentence routing use different budgets.}
    \label{tab:appendix-redacbench-operating-points}
\end{table}

\begin{figure}[htbp]
    \centering
    \includegraphics[width=\linewidth]{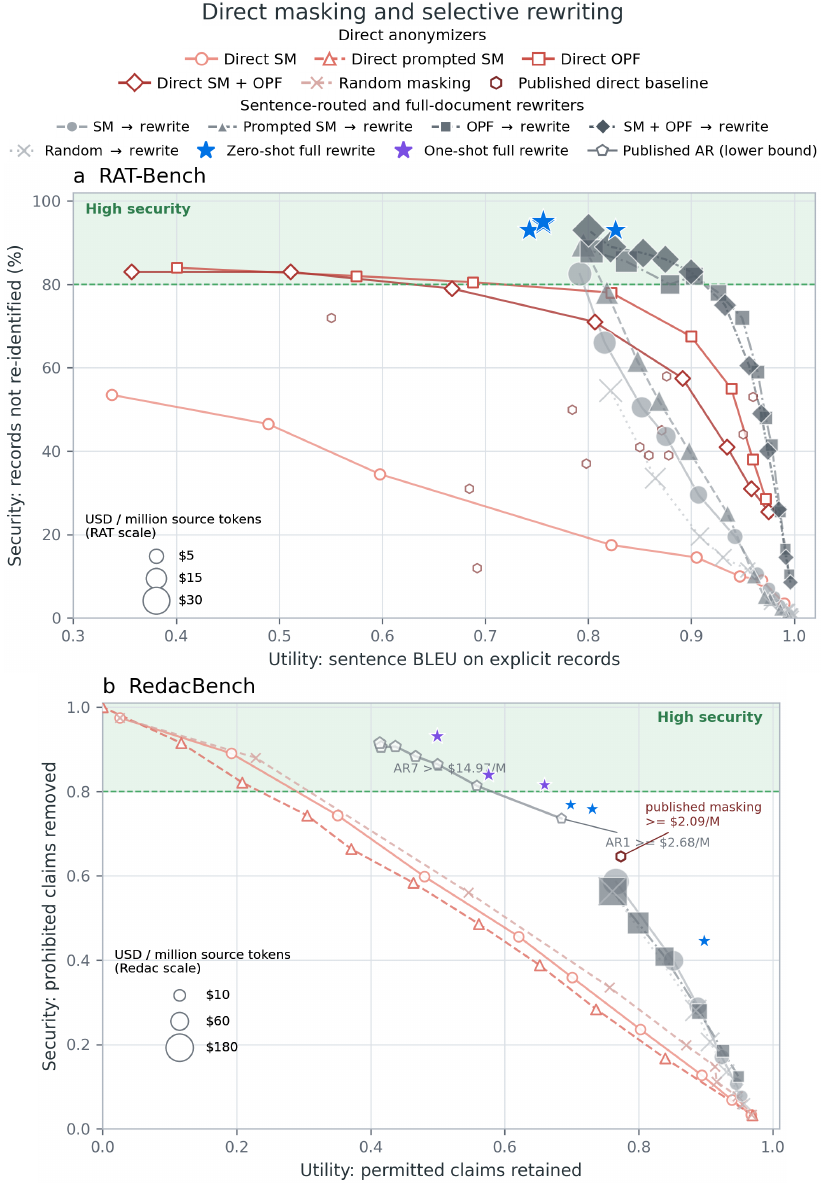}
    \caption{Security, utility, and cost on RAT-Bench and RedacBench.
    RAT-Bench security is one minus identity recovery rate; RedacBench
    security is the fraction of prohibited propositions removed. Bubble
    area shows USD per million source tokens, with independent scales for
    the two benchmarks. Published RedacBench costs are lower bounds;
    published RAT-Bench costs are unavailable.}
    \label{fig:anonymization}
\end{figure}

\paragraph{Cost accounting.}
Costs are normalized by source tokens in the evaluated documents: 266,247
for 200 RAT-Bench transcripts and 135,404 for 513 RedacBench documents.

For recorded rewriter input, cached input, and output counts
$N_{\mathrm{in}}$, $N_{\mathrm{cached}}$, and $N_{\mathrm{out}}$, prices
$p_{\mathrm{in}}$, $p_{\mathrm{cached}}$, and $p_{\mathrm{out}}$ in dollars
per token, and estimated selector cost $C_{\mathrm{selector}}$, we report
\[
C_{\mathrm{USD/M}}
=
\frac{10^6}{N_{\mathrm{source}}}
\left[
p_{\mathrm{in}}\bigl(N_{\mathrm{in}}-N_{\mathrm{cached}}\bigr)
+
p_{\mathrm{cached}}N_{\mathrm{cached}}
+
p_{\mathrm{out}}N_{\mathrm{out}}
+
C_{\mathrm{selector}}
\right].
\]

The assumed prices per million uncached input, cached input, and output
tokens are \$5.00, \$0.50, and \$30.00 for GPT-5.5, and \$0.40, \$0.10,
and \$1.60 for published GPT-4.1-mini methods. Selector estimates assume
a warm H200 at \$5 per hour. Costs include rewriter policy and context
tokens and exclude startup, attacker, and evaluator costs. GPT-5.5 costs
use recorded cache usage. Published GPT-4.1-mini costs are lower bounds
because intermediate traces are unavailable.

At 20\% RedacBench routing, OPF makes 1,402 calls across 513 documents,
using 789,669 input and 56,832 output tokens. Repeated policy and context
raise the cost to \$41.97 per million source tokens, above \$37.53 for a
full document rewrite.

\section{Generation-Probability Bound for Global Surprise Masking}
\label{sec:global-sm-probability}

Global SM masks content that is unlikely under the reference model. Here we quantify how likely a span must be to remain fully unmasked and bound the expected number of samples needed to generate it.
In particular, we consider just the global rule in Equation~\ref{eq:window-surprise}. Local ranks, contrastive scores, and consensus masking do not generally satisfy these assumptions. Exact duplicate propagation preserves the bound when applied after the global rule because it only adds masked positions.

\subsection{Probability bound}

The following result applies only to the global length-$n$ window rule in
Equation~\ref{eq:window-surprise}. Assume every length-$n$ window fully
contained in the span is scored, token log probabilities are finite, and a
window at or above the surprise threshold masks all tokens it covers.

\paragraph{Proposition.}
Let $w=(x_s,\ldots,x_{s+m-1})$ be an $m$-token span, with $m\geq n$, whose
tokens are all unmasked at threshold $\tau$. Then
\[
p_{\mathcal M}(w\mid x_{<s})
\geq
\exp\!\left(-n\tau\left\lceil\frac{m}{n}\right\rceil\right).
\]
If independent $m$-token continuations are repeatedly sampled from the same
reference-model conditional distribution, the expected number of samples
until one equals $w$ is at most
\[
\exp\!\left(n\tau\left\lceil\frac{m}{n}\right\rceil\right).
\]

\paragraph{Proof.}
For a length-$n$ window beginning at $i$, write
\[
\Lambda_i
=
\sum_{r=0}^{n-1}
\log p_{\mathcal M}(x_{i+r}\mid x_{<i+r}).
\]
No window fully contained in $w$ can have been selected; otherwise its tokens
would have been masked. Therefore $\Lambda_i\geq-n\tau$ for every such window.
The weak inequality also covers any implementation convention that selects
threshold ties.

Set $q=\lceil m/n\rceil$. Choose windows starting at
$s,s+n,\ldots,s+(q-2)n$, with the final window starting at $s+m-n$. These
$q$ windows cover every token in $w$ at least once. Because each token log
probability is nonpositive, counting overlapping tokens again can only lower
the sum:
\[
-qn\tau
\leq
\sum_{j=1}^{q}\Lambda_{a_j}
\leq
\sum_{r=0}^{m-1}
\log p_{\mathcal M}(x_{s+r}\mid x_{<s+r})
=
\log p_{\mathcal M}(w\mid x_{<s}).
\]

Exponentiating gives the probability bound. Independent repeated samples have a geometric waiting time with success probability $p_{\mathcal M}(w\mid x_{<s})$, giving the bound on its expectation.

\subsection{Temperature and decoding}

The sampling bound assumes temperature $1$ and no truncation. We next examine how changing the decoding rule changes the probability of generating the same unmasked span.

For $T>0$, temperature sampling uses
\[
q_T(v\mid h)
=
\frac{p_{\mathcal M}(v\mid h)^{1/T}}
     {Z_T(h)},
\qquad
Z_T(h)=\sum_{u}p_{\mathcal M}(u\mid h)^{1/T}.
\]
For the target span $w$,
\[
\log q_T(w\mid x_{<s})
=
\frac{1}{T}\log p_{\mathcal M}(w\mid x_{<s})
-
\sum_{i=0}^{m-1}\log Z_T(x_{<s+i}),
\]
where each history includes the target tokens already generated. The normalizers depend on the full token distribution at each step, so the original span probability alone does not determine its probability at another temperature.

For $T<1$, a target near the top of the token distribution can become more likely, while a less likely target can become less likely. As $T\to0$, the span probability approaches one if every target token is the unique maximum at its step, and zero if any target token is not a maximum. Tied maxima share the limiting probability mass. Top $p$ truncation can instead assign zero probability to a required token by excluding it from the nucleus. Our Code PII extraction experiments use temperature $1$ and top $p=1$.

\end{document}